\documentclass[a4paper,fleqn,usenatbib]{mnras}

\usepackage[T1]{fontenc}
\usepackage{ae,aecompl}

\usepackage{longtable}

\usepackage{graphicx}	
\usepackage{amsmath}	
\usepackage{amssymb}	
\usepackage{lscape}

\usepackage{subfigure}

\title[]{UV-FIR SED modeling of AGN in IR-luminous galaxies up to z$\sim$2.5:     Understanding the effects of torus models }

\author[A. Sokol et al.]{Alyssa~D. Sokol$^{1}$\thanks{E-mail: asokol@umass.edu},
M.Yun $^{1}$,
A. Pope$^{1}$,
A. Kirkpatrick$^{2}$,
K. Cooke$^{3}$
\\
$^{1}$Department of Astronomy, University of Massachusetts, Amherst, MA, USA \\
$^{2}$Department of Physics and Astronomy, University of Kansas, Lawrence, KS, USA\\
$^{3}$Association of Public and Land-grant Universities, 1220 L St NW Suite 1000, Washington, DC 20005
}

\pubyear{2017}

\begin{document}
\label{firstpage}
\pagerange{\pageref{firstpage}--\pageref{lastpage}}
\maketitle
\begin{abstract}
UV-FIR SED modeling is an effective way to disentangle emission between star formation (SF) and active galactic nuclei (AGN) in galaxies; however, this approach becomes uncertain for composite AGN/SF galaxies that comprise 50-70\% of IR-samples.  Cosmic X-ray background (XRB) models require a large fraction of  obscured AGN to reproduce the observed XRB peak, motivating reliable SED analyses in objects where the AGN may be ``buried" in the galaxy and in the mid-IR to far-IR SED.  In this paper, we study a 24$\mu$m-selected ($S_{24}$ > 100$\mu$Jy) sample of 95 galaxies with $0 \% < f_{MIR,AGN} < 100 \%$, 0.4 < z < 2.7, and $10^{11}$L$_{\odot}$ < L$_{IR}$ < $10^{13}$L$_{\odot}$. We test the performance of AGN models ranging in torus optical depth via SED fitting, comparing results with Spitzer MIR spectroscopy and X-ray observations. Best-fit torus optical depth can shed light on whether these galaxies host a luminous obscured AGN population. We find that permitting a broader AGN SED parameter space results in improved fit quality with higher optical depths, higher FIR AGN contributions, and higher $L_{Bol}$, impacting the bright-end of the $L_{Bol}$ luminosity function. Our results suggest there may be a population of dust-obscured composites that are bolometrically significant but have their AGN mostly hidden in the mid-IR SED. If so,  literature applications of SED fitting that often simplify AGN models or omit optically thick torrii may largely underestimate AGN contribution from composite sources, as these sources are both numerous and have solutions sensitive to the assumed range of AGN models.
\end{abstract}




\section{Introduction}

It is thought that nearly every massive galaxy hosts a super-massive black hole (SMBH) on the order of $10^{6}-10^{9} M_{\odot}$ at its center; however, the connection between star formation and the fueling of this nuclear activity remains an outstanding issue in galaxy evolution \citep{Kormendy1995,Alexander2012,Harrison2017}.  Observational studies have shown that the black hole accretion rate density (BHARD) and star formation rate density (SFRD) both peak around z$\sim$ 1-2 and decline to the present day \citep{Hopkins2006,Silverman2009,Aird2010,Aird2015,Delvecchio2014}. To understand the nature, if any, of this connection, it is necessary to separate the contribution between black hole accretion and star formation (SF) within the same galaxies for a large sample over a significant range of redshifts and luminosities. 

While optical diagnostics offer some distinction between SF and nuclear accretion processes, they are not ideal for dust obscured objects and composite AGN/SF galaxies \citep{Baldwin1981,Kauffmann2003, Trump2015}. The optimal selection wavelength to probe this mass buildup is in rest-frame mid-infrared (MIR; 5-12$\mu$m) and infrared (IR;  8-1000$\mu$m).  In this regime, heated dust emission from both SF and AGN coincide to produce unique spectral energy distribution (SED) signatures that allow us to study both star formation and black hole accretion out to  z $\sim$ 2.5 \citep{DaleHelou2002,Ciesla2015,Assef2018}. We note that while there are inconsistent definitions of the IR regime across the literature, in this paper we define both the total IR and FIR to be in the same wavelength range of 8-1000$\mu$m.

In regular star-forming galaxies (SFGs), emission from older stellar populations steadily rises in the optical and peaks near rest-frame 1$\mu$m, dropping off to exhibit a dip in their SEDs between $\sim$ 1-5$\mu$m\citep{LiDraine2001,DraineLi2007}.  At $\lambda$ > 5$\mu$m, SFG spectra show a series of polycyclic aromatic hydrocarbon (PAH) emission features that arise from the heating of the interstellar medium (ISM) by recent star formation in the rest-frame $\sim$ 5-12$\mu$m range \citep{Pope2008,Sajina2012}.  

As for AGN, when a galaxy's central supermassive black hole (SMBH) is actively accreting material,  nuclear activity heats the surrounding dust to produce an additional mid-IR emission component \citep{Ciesla2015,Malek2017}. In the rest-frame near-infrared (NIR) to MIR, $\sim$ 1-15$\mu$m, both unobscured and obscured AGN are expected to have warm dust power-law SEDs; this feature is produced when nuclear photons are absorbed by the surrounding dusty torus and re-radiated into longer wavelengths \citep{Elvis1994, Mullaney2011}. This warm dust emission from the AGN  will fill in the notorious 1-5$\mu$m SFG dip and saturate the 5-12$\mu$m PAH emission features, producing a prominent rising MIR SED indicative of AGN activity  \citep{Donley2008, Donley2012}.


To disentangle observed SEDs into their respective AGN and SF components, UV-FIR SED fitting can be employed with sufficient multi-wavelength broadband photometry and known source redshift. SED decomposition techniques apply a simultaneous three component fit to the SED  and often adopt energy balance accounting for (1) UV/optical emission from starlight, (2) a MIR-FIR AGN emission component, and (3) FIR emission from dust-reprocessed SF. Each of the three components require an assumed SED library to model emission respectively. Some widely adopted SED fitting codes include SED3FIT \citep{Berta2013}, CIGALE, and MAGPHYS+AGN \citep{Chang2017}. Many works have applied these methods to compute AGN and SF luminosities for galaxy populations \citep{Delvecchio2014,Delvecchio2017,Chang2017}; however, the accuracy and/or systematics of these methods for weakly contributing AGN are less addressed. 

A key motivation behind AGN decomposition, and understanding the reliability of SED fitting techniques,  is to identify and characterize elusive AGN populations that are not overwhelmed by AGN emission ($ 0 \% < f_{MIR,AGN} < 50\%$). These sources can be characterized more effectively with MIR spectroscopy revealing subtle changes in $f_{MIR,AGN}$ (see \cite{Sajina2022} for a recent review). However, this rich data is only available for a few hundred sources, making it difficult to track statistically significant samples out to high redshfit \citep{Pope2008,Kirkpatrick2012,Sajina2012}.

Elusive populations not overwhelmed by AGN emission may also be deeply buried, obscured AGN contributing to the cosmic X-ray background \citep{Ueda2003, Gilli2007, Daddi2007,Treister2009, Comastri2015}. These Compton Thick ( $N_{H} > 10^{24}$ cm$^{-2}$ ) objects can be missed in X-ray surveys, as extremely high columns of dust can absorb even the hardest X-ray photons \citep{Fiore2009,Lansbury2015,Hickox2018}. While MIR and IR AGN studies are more likely to detect heavily obscured objects missed in X-ray, the issue remains that deeply buried AGN may have SEDs indistinguishable from SFGs \citep{Ricci2015,Lambrides2020}.  For this reason, it is important to closely study AGN candidates that are less pronounced in the MIR to investigate and characterize their plausible bolometric output.  

The picture of broadband SED fitting is further complicated when we also consider the predicted variability in shape of the underlying MIR-FIR AGN SED.  AGN SED characteristics are determined by the geometry of dust surrounding the central SMBH; circumnuclear dust is commonly modeled as a torus shape, though the distribution of dust within the torus as smooth versus clumpy has been widely debated \citep{Nenkova2008,Hatz2010,Fritz2006,Feltre2012}.   According to smooth radiative transfer models \citep{Fritz2006,Feltre2012}, the AGN MIR and FIR SED shape  is affected significantly by the presence of a Silicate 9.7$\mu$m absorption feature.  Si 9.7$\mu$m absorption strengthens when toroidal dust becomes more apparent along the line of sight, predicted for edge-on systems (Type-2 AGN) and/or when the torus is optically thick. Conversely, the Si 9.7$\mu$m feature appears in emission when systems are observed face-on or the central SMBH is not heavily dust-enshrouded \citep{Hao2005,Sieben2005}. Torus optical depth, $\tau_{9.7}$, also determines the strength of AGN-related FIR emission, where the dust from optically-reddened AGN SEDs will be absorbed and reprocessed to produce significant AGN FIR emission.

While a variety of AGN SED models exist,  there is no clear consensus on which models, or their corresponding NIR, MIR, and FIR parameter spaces, are most accurate to describe the plethora of multi-wavelength AGN observations. Despite the lack of consensus, users applying SED fitting are confronted with the decision of which AGN models to employ in their analyses. While some argue for adopting a simpler AGN SED library, \cite{Gonzalez2019b} find that the largest SED libraries, upwards of 132,000 SEDs, provide better fits to MIR spectra, supporting the use of well-sampled SED libraries. However, applying a large volume of varied AGN models to broadband SED fitting codes is computationally expensive, especially for large samples. Further, an excess of available free parameters combined with limited broadband SED coverage can lead to degenerate fits for large libraries. As a result, many works instead apply smaller sets (typically 4 - 10) of averaged AGN models \citep{Chang2017}, empirical or theoretical, to derive AGN parameters.

Despite the computational advantage of working with a smaller, restricted set of AGN models, simplifying a set of AGN models requires assumptions that can contribute to varied results or larger uncertainties. For example, large torrii radii were initially considered unrealistic \citep{Pier1992} under the assumption that FIR  emission is dominated by star-formation rather than AGN.  However, the relationship between torus size, smoothness, and FIR emission is not well-understood, and applying cuts to large AGN torus libraries places restrictions on physical torus properties that may be incorrect. Additionally, it is unclear whether templates adopted from local AGN that are more gas rich \citep{Tacconi2010} can be confidently applied to sources at higher redshifts. Thus, studying the redshift evolution of AGN properties may be hindered by the use of AGN templates constructed from local samples.  

\cite{Henry2019} find that adequate mid-IR coverage is crucial to properly tease out AGN contribution to the mid-IR SED. While the community acknowledges that the lack of AGN model constraints poses a problem and can lead to fit degeneracies \citep{Alberts2020}, the effects of potential systematics on additional AGN parameters are not widely addressed.  For these reasons, exploring results from both an expansive AGN SED parameter space and a restricted set can provide insight on how torus model simplifications and sparse MIR coverage, may impact or bias SED decomposition results.


 In this work, we apply SED fitting to a 24-$\mu$m selected sample of 95 galaxies with $0.4 < z < 2.7$,  $0 < f_{MIR,AGN} < 100\%$, and Spitzer/IRS spectroscopy.  $f_{MIR,AGN}$ values are computed in \cite{Kirkpatrick2012} using MIR Spitzer spectroscopy, though we also evaluate how this MIR AGN contribution compares when derived with SED fitting. We analyze the impact of AGN SED models on best-fit AGN MIR and FIR fractional contributions, AGN bolometric luminosities, AGN torus optical depth $\tau_{9.7}$ , and SFR. Such parameters will affect the accuracy of AGN bolometric luminosity functions (LFs), particularly for composite sources with an ambiguous mixture of SF and AGN activity.   In their latest review of AGN traced by Spitzer Space Telescope, \cite{Lacy2020} support the notion that a complete census of supermassive black hole growth should not only count AGN-dominated sources, but also include these intermediate composite sources. A detailed understanding of both sets of LFs is crucial to understand the simultaneous universal mass assembly via supermassive black hole accretion and star formation. 
 
The paper is structured as follows.  In Section~\ref{sec:sample} we introduce our 24$\mu$m-selected sample and its multi-wavelength data used for SED fitting.  In Section~\ref{sec:torus} we describe the UV-FIR SED fitting code, SED3FIT \citep{Berta2013}, used to disentangle AGN and SF emission. We also summarize the current state of diverse AGN torus models and describe the models we implement and test throughout this paper.   In Section~\ref{sec:results} we present our main results by comparing SED fitting quantities between runs using a full and limited AGN torus library. We analyze the effects of torus models on MIR AGN fraction, $L_{Bol}$, and SFRs, sensitive to AGN contamination when derived in the FIR.   In Section~\ref{sec:discussion} we discuss the  broader implications of AGN torus model choice;  we describe how AGN model sensitivities may impact estimates of the AGN incidence rate in IR-samples, AGN bolometric luminosity functions, and our ability to detect and characterize AGN emission in heavily dust obscured sources missed by X-ray surveys.   Finally, in Section~\ref{sec:summary} we summarize our key results. Throughout the paper we assume a $\Lambda$CDM cosmology with parameters $H_{0} = 70$ km s$^{-1}$Mpc$^{-1}$, $\Omega_{m} = $0.3, and $\Omega_{\Lambda}$=0.7 \citep{Spergel2007}.

\section{Sample Description}
\label{sec:sample}
\begin{figure*}
\centering
\includegraphics[width=\textwidth]{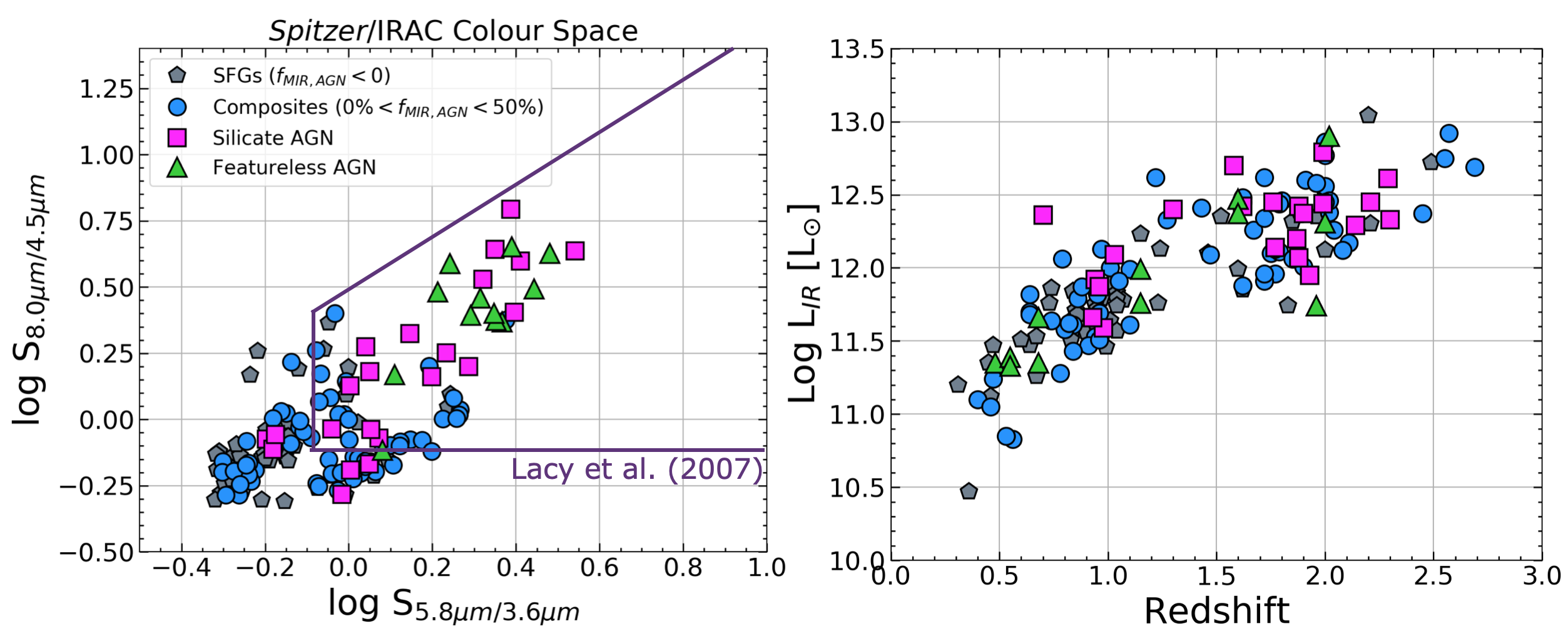}
\caption[3]{\textit{Left:} IRAC Mid-IR Color plot for the 151 sources in the \cite{Kirkpatrick2012} sample spanning 0.4 $ < z <$ 2.7. The sample contains 70 sources from GOODS-N and 81 sources from ECDFs selected to have $S_{24}>100$ $\mu$ Jy and Spitzer/IRS MIR spectroscopic observations. Of the 151 sources, 34 are spectroscopically classified as either Silicate AGN (magenta squares)  or Featureless AGN (green triangles), 64 are composite AGN-SF galaxies (blue circles), and 49 are star-forming galaxies (gray pentagons). For reference we show the IRAC AGN color selection wedge from \cite{Lacy2007}, but no color selection is applied to the original sample selection in \cite{Kirkpatrick2012} or this work. \textit{Right:} Total IR Luminosity (rest-frame 8 - 1000$\mu$m) as a function of redshift for all 151 sources in the sample.}
\label{fig:colorplot}
\end{figure*}

 The objective of this work is to perform multi-wavelength (UV-FIR) AGN SED decomposition for AGN  ranging in MIR strength to illustrate how results are driven by the predicted variation in underlying AGN SED models.  To add meaningful reference to our exploration of these biases,  we use a sample of 24$\mu$m-selected  ($S_{24}>100 \mu$Jy) sources at $0.4 < z < 2.7$ in the GOODS-N and ECDFS fields studied by \cite{Kirkpatrick2012}. This $S_{24}$ detection threshold is chosen to deliberately study MIR-bright sources observable with Spitzer/IRS. We describe below the full sample studied in \cite{Kirkpatrick2012}, consisting of 151 sources with $10^{11}$ L$_{\odot}$ < $L_{IR}$ < 10$^{13}$ L$_{\odot}$. 
 

\cite{Kirkpatrick2012} derive MIR AGN fractions for all 151 sources via spectral decomposition (see Section~\ref{subsubsec:spec_decomp}).  They divide the sample by MIR AGN fraction into four groups : Silicate AGN, Featureless AGN, Composites, and SFGs. The two AGN groups are differentiated by the presence or absence of a 9.7$\mu$m  Silicate absorption feature in MIR spectroscopy. This feature, caused by circumnuclear dust absorption along the line of sight, is visually apparent in spectra of  Silicate AGN and treated in spectral fitting (see again Section~\ref{subsubsec:spec_decomp}). Featureless AGN lack this absorption feature,  exhibiting instead a rising MIR SED slope indicative of minimal nuclear dust absorption along the line of sight. Of the 151 sources total, 34 are AGN with $f_{MIR,AGN} > 50\%$; 22 of the 34 AGN are classified as Silicate AGN and 12 are classified as Featureless AGN. There are 4 additional AGN sources in the parent sample of 151 that did not have adequate spectroscopy to be classified as Silicate or Featureless; we omit these sources from this study. We also omit from this study 3 AGN sources with no broadband 24$\mu$m data, leaving a total of 31 classified AGN to be analyzed in this work. Composites contain a mixture of AGN and SF emission but are not dominated by the AGN component; 64 sources are classified as composites, with $0 < f_{MIR,AGN} < 50\%$. SFGs are galaxies that are estimated to have no AGN contribution to their MIR spectra;  49 sources are classified as regular star forming galaxies with $f_{MIR,AGN} \sim  0\%$. 

Rather than a single flux limit selection, this sample requires both $S_{24}>100 \mu Jy$  and Spitzer/IRS spectroscopic observations. Despite the additional spectroscopic requirement, the 151-source sample is roughly consistent with the full Spitzer/MIPS 24$\mu$m population above the flux limit $S_{24}>100 \mu$Jy. \cite{Kirkpatrick2012} test this, finding a similar AGN incidence fraction, $25\%$, and AGN $S_{24}$ flux distribution between the parent flux-limited population  and the narrowed-down selection. This sample is also  representative of an IR-selected sample; 70\% of sources have a Herschel/PACs detection at 100$\mu$m and 67\% of PACs sources in GOODS-N have $S_{24}>100 \mu$Jy. Photometric fluxes and uncertainties from Spitzer and Herschel are given in \cite{Kirkpatrick2012}, along with spectroscopic redshifts for the entire sample derived using Spitzer/IRS spectroscopy; these data are used to perform SED decomposition for this work.

In the left panel of  Figure~\ref{fig:colorplot} we show the IRAC color distribution of all 151 sources in this sample along with selection wedge from \cite{Lacy2007} for reference. While we do not adopt any MIR color selection wedge, the color plot and wedge shown illustrate that composite sources commonly reside outside of MIR selection criteria and overlap with SFGs.  On the right panel of Figure~\ref{fig:colorplot} we show the IR Luminosity of the sample as a function of redshift. To examine AGN contribution systematics, we use only those with at least a 24$\mu$m detection and a spectroscopically computed MIR AGN fraction $> 0\%$, leaving a total of 95 sources with $ 0\% < f_{MIR,AGN} < 100\%$ for analysis. 31 of 95 sources are AGN with $f_{MIR,AGN} > 50\%$ (22 Silicate AGN and 12 Featureless AGN). 64 of 95 sources are composites, with $ 0\% < f_{MIR,AGN} < 50\%$. Of the 95 total sources, 65 have detections at both 16$\mu$m and 24$\mu$m and 30 only have a 24$\mu$m MIR detection.

We choose this sample for a few reasons: (1) These sources have Spitzer/IRS MIR spectroscopy, allowing us to estimate a calibration between the two decomposition methods, and (2) The sample can serve as an analogue to larger samples that require analysis via AGN SED decomposition. With spectroscopic data to support a non-negligible AGN presence in the 64 composites, we have the opportunity to explore the performance of broadband SED fiting for sources not overwhelmed by AGN emission in the mid-IR.

\begin{figure*}
\centering
\includegraphics[width=\textwidth]{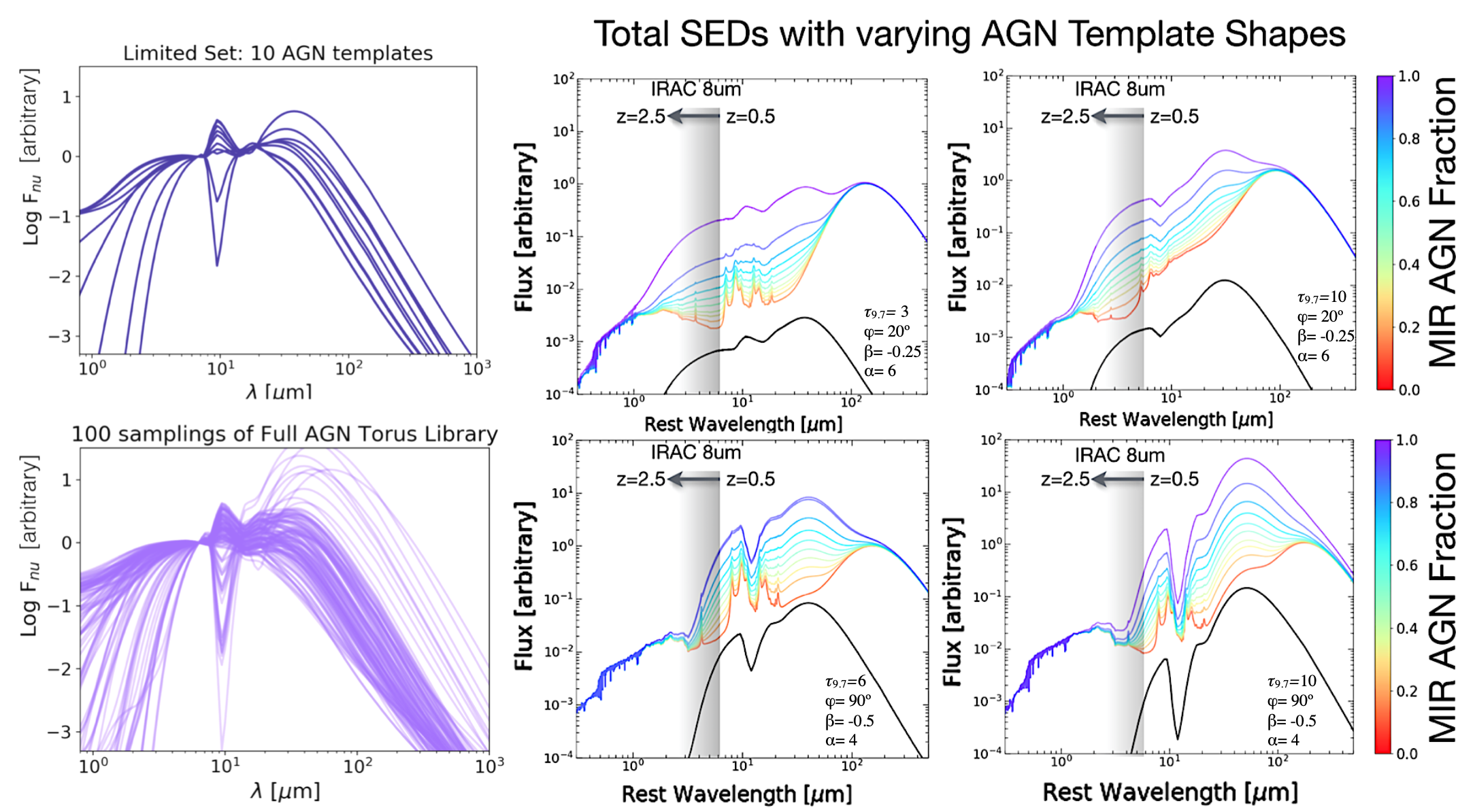}
\caption[2]{ \textit{Left column:} Two AGN theoretical template selections \citep{Fritz2006,Feltre2012} used for SED decomposition testing in this work and normalized at 7$\mu$m for ease of viewing. To test the sensitivity of allowed AGN models, we apply both a restricted set and larger library sampling for comparisons throughout the paper. The bottom figure shows 100 random samplings of the full AGN torus library containing 21000 models total. The top figure shows the limited set of  10 default AGN models included with the SED Decomposition code download \citep{Berta2013}. Primary differences between each AGN model choice are for torus optical depth and viewing angle, where sampling the full library includes more viewing angles and higher optical depths. As a result, this model selection includes AGN templates with deeper Si absorption features and more extended FIR emission. See Table~\ref{table:agnparams} for a detailed comparison of parameter differences for each model setting. \textit{Right:} A demonstration of how four different underlying AGN models uniquely affect their galaxy total SED shape at NIR-IR wavelengths. Each panel shows 10 total SEDs for 10 MIR AGN fractions produced by changing the normalization of a single underlying AGN template. AGN SEDs are shown in black at an arbitrary normalization to illustrate the underlying shape and the colorbar represents the rest-frame 5-12$\mu$m AGN contribution for each total SED. The shaded band represents the Spitzer/IRAC 8$\mu$m coverage for $0.5 < z < 2.5$. When the AGN SED is more reddened, in the bottom two figures, as redshift increases the Spitzer 8$\mu$m band no longer traces varying MIR AGN contributions as it does in the top two figures.  }
\label{fig:torusmodels}
\end{figure*}


\section{  SED Fitting and AGN Torus Models}
\label{sec:torus}

\subsection{Broadband SED Decomposition: SED3FIT Code}
\label{subsec:sed3fit}

 To perform SED fitting we have adopted the publicly available code, SED3FIT \footnote{SED3FIT code here: http://cosmos.astro.caltech.edu/page/other-tools} \citep{Berta2013}.  SED3FIT is a three-component SED fitting routine that accounts for contribution from optical starlight, reprocessed dust emission from star formation \citep{dacunha2008}, and AGN accretion disk and dusty torus \citep{Fritz2006, Feltre2012}.  Stellar and dust emission are linked by energy balance, such that energy absorbed by dust at UV-optical wavelengths is re-emitted at infrared wavelengths. The code requires input for broadband photometric data, photometric error, and redshift, resulting in a simultaneous three-component fit. For this work, these values are adopted from \cite{Kirkpatrick2012} (see Section~\ref{sec:sample}). For each best-fit parameter, a corresponding probability distribution function (PDF) is available, allowing for confidence ranges on parameter estimates to be well-understood by the user. Several studies have used SED3FIT to decompose SEDs of AGN selected at IR \citep{Delvecchio2014} and radio \citep{Delvecchio2017, Delvecchio2018} wavelengths. 
 
 Important best-fit quantities returned by SED3FIT include AGN bolometric luminosity, AGN fractions at various ranges, IR luminosity from dusty star formation, and an array of best-fit parameters that correspond to the chosen AGN torus model.  Further, since the entire UV-FIR best-fit SED is returned with its components, we can compute specific quantities by integrating models over desired wavelength ranges.  SED3FIT allows the user to input any desired AGN torus models, though it includes a set of theoretical models from \cite{Fritz2006} and \cite{Feltre2012}.   With the advantage of the MIT Supercloud supercomputing facility, we adopt an expansive AGN template parameter space to disentangle broadband SEDs of these objects, gaining a deeper understanding of systematics affecting the decomposition of weakly contributing AGN or composites. The models tested in this work are described in Section~\ref{subsec:mymodels}, however, we discuss the plethora of AGN SED models available and their physical differences below.

 \subsection{Clumpy vs. Smooth AGN Torus Models} 
 \label{subsec:clumpysmooth}
 
Due to the elusive multi-wavelength nature of AGN and the difficulty of directly observing these $\sim$ 5 pc wide nuclear structures \citep{Ramos2017},  the distribution of dust surrounding AGN accretion disks has led to a variety of proposed physical models. Model IR SEDs of AGN tori derive their shape from dust geometry and their power from the amount of energy absorbed from the nuclear region \citep{Netzer2015}.  The earliest models propose a simple, smooth and homogenous toroidal disk \citep{Pier1992,Ef1995a,Fritz2006}. Smooth models define circumnuclear dust as a continuous distribution in a toroidal  or flared disk shape in accordance with the unified scheme for AGN \citep{Antonucci1993}, where differences in AGN signatures may be attributable to different viewing angles.  Early smooth toroidal disk models have been challenged by observations; if accurate, observational signatures should correlate with inclination angle such that edge-on Type-2 AGN show Silicate absorption that does not appear in  face-on Type-1 objects. These models could not explain observed Silicate absorption features in some Type-1 objects \citep{Roche1991} or MIR radiation from the polar ionization cone \citep{Honig2013}, where the line of sight is not heavily dust-obscured. 
 
 To mitigate possible inconsistencies, a clumpy torus model was later proposed \citep{Nenkova2008,Honig2013}. The toroidal model of \cite{Nenkova2008} uses standard ISM composition to describe dust grain chemistry  and characterizes the torus as a concentrations of dust clumps or clouds. The clumpy model of \cite{Honig2010} is similar, with a volume filling factor between clumps of high optical depth. The filling factor describes how tightly packed clumps are within the torus, where a model with an extremely high filling factor is physically similar to an optically thick smooth torus model. Two-phase models also successfully reproduce observed AGN MIR emission. These models combine the aforementioned features, allowing a dusty homogenous disk, clumpy medium, or both \citep{Sieben2005,Feltre2012,Sieben2015}. Similarly to a homogenous disk model,  increasingly edge-on viewing angle results in a reddened optical slope, Silicate absorption feature, and enhanced FIR emission.  Clumpy models and two phase models are differentiated by the allowed filling factor between clumps. 
 
Though the clumpy and two-phase models may be more physically feasible than a continuous disk, smooth models are simpler computationally and can provide reasonable estimates of the IR AGN SED.  Further, clumpy models also face challenges matching observations to short wavelength hot-dust emission from face-on AGN \citep{Feltre2012}. Evidently, due to conflicting models and their ability to describe the full AGN population, there is no clear consensus on appropriate AGN SED models and their corresponding dust distribution.  Since studies remain inconclusive, we deliberately choose to work with a large set of smooth torus models that vary significantly in shape to address the full suite of possible AGN torus configurations.

 

 

\begin{table*}
\begin{tabular}{ |l |l |l | l| }

Parameter  & \textbf{Full Library (21000 Models)}   & Limited Library (10 Models)  &  Description\\

\hline
\hline
$R_{max}/R_{min}$ & \textbf{10,30,60,100,150}  & 30  &   Ratio of Maximum to minimum dust torus radius\\   
 
\hline
\hline

$\Theta$ & \textbf{60,100,140}& 100&  Opening Angle of Torus \\  

\hline
\hline

$\tau_{9.7}$ & \textbf{0.1, 0.3, 0.6 , 1.0, 3.0, 6.0, 10.0 } & 0.1, 0.3, 1.0, 3.0, 6.0 &  Torus Optical Depth \\  

  \hline
  \hline
  
 $ \beta$  & \textbf{-1.0, -0.75, -0.50, -0.25, 0.0}   & 0.0&  Radial dust distribution in torus\\  

  \hline
  \hline

  $\gamma$  & \textbf{0.0, 2.0, 4.0, 6.0}   & 0.0 &  Polar dust distribution in torus\\  

      \hline
  \hline
  
    $\phi$  & \textbf{0, 10, 20, 30, 40, 50, 60, 70, 80, 90}  & 0, 90  &Viewing Angle (0=Face on, 90=Edge-on)\\

\end{tabular}
\caption[2]{Parameters for two samplings of theoretical AGN models in \cite{Fritz2006} and \cite{Feltre2012} applied in this work. The full AGN library contains 21000 Models. In this paper we fit objects twice: (i) Using 100 random samplings of the full 21000-model AGN library (2nd column) and (ii) Using a discrete set of 10 models (3rd column) without random sampling. The limited library distributes the 5 possible $\tau_{9.7}$ values amongst 5 edge-on models and 5 face-on models, yielding a total of 10 possible combinations.}
\label{table:agnparams}
\end{table*}

\subsection{Models used in this paper}
\label{subsec:mymodels}


For our AGN SED decomposition we utilize the smooth radiative transfer toroidal models of \cite{Fritz2006} and \cite{Feltre2012}. This library includes 21000 models that have each been analyzed for 10 different lines of sight, $\phi$, uniformly distributed from edge-on to face-on.  Each AGN template includes emission from the hot accretion disk in addition to reprocessed emission from the surrounding dusty torus. The central heating source is modelled with a broken 3-part power law, while at longer wavelengths the radiation scales as a Planck spectrum in the Rayleigh-Jeans regime. Six different total parameters are used to identify each torus model: $R_{max}/R_{min}$ , the ratio of outer-to-inner radius; $\Theta$, the opening angle; $\tau_{9.7}$, the equatorial optical depth at 9.7$\mu$m, $\beta$ and $\gamma$, the radial and height slopes of the density profile; and $\phi$, the line-of-sight viewing angle. To robustly explore the consequences, if any, of choice of models, we explore this grander suite of models in our analysis at various settings.

We adopt two versions of sampled AGN template space for this work: (1) A limited set of 10 AGN models, hereafter, the limited AGN library (2) 100 random samplings of the entire 21000-model library, hereafter, the full AGN library. Each of the optical and IR libraries contain 50,000 models and all runs used in this study adopt 100 random samplings of both for fitting. Both sets of allowed models are shown in the left column of  Figure~\ref{fig:torusmodels}, where they are normalized at 7$\mu$m for ease of viewing comparison. In Table~\ref{table:agnparams} we show the different parameter values, described above, present in each model setting.

The key parameters restricted in the limited set of 10 models are viewing angle and torus optical depth.  Sampling the full library allows 10 intermediate viewing angles to be tested, compared with the limited set that represents only 2 (edge-on and face-on) viewing angles. Visually, the models roughly trace the same parameter space with the exception of the Silicate depth and FIR emission strength.  Physically, this difference can be described by the addition of higher optical depth AGN torus models. In the full library, this value extends to $\tau_{9.7} = 10$, where the highest $\tau_{9.7}$ value sampled by the limited set is $\tau_{9.7} =6$.  The increase in optical depth simultaneously increases the far-IR AGN emission.




In Section~\ref{sec:sample}, we presented the sample selection and defined its MIR AGN fraction classification. While this quantity is often used as a proxy for AGN dominance, Figure~\ref{fig:torusmodels} illustrates that similar MIR AGN fractions can produce a range of SED shapes that depend on the underlying AGN model.   The top two panels show an underlying AGN SED that is optically bright with no Silicate absorption, consistent with a face-on viewing angle.  The bottom two panels show stronger Si absorption with a reddened optical slope and enhanced FIR emission. The figure demonstrates how the MIR region may exhibit drastically different features depending on the underlying AGN SED characteristics. 

We also show the redshifted rest-frame location of the longest wavelength Spitzer/IRAC data point, 8.0 $\mu$m. In the bottom two panels, for an obscured torus the observed 8.0 $\mu$m band is insensitive to changes in AGN contribution shortly above $z \gtrsim 1$.  As a result of this unaffected MIR SED, MIR color will also not correlate with AGN contribution and can be confused when AGN model variance is taken into account.  Additionally, fitting may be affected by the proximity of constraints near the 9.7$\mu$m Si absorption feature.  This implies that SED decomposition routines are likely sensitive to MIR data beyond the IRAC bands to properly distinguish heavily reddened AGN SED models. 

\section{Results }
\label{sec:results}



 
We explore the effects of AGN torus models on SED decomposition by comparing results between fits using a restricted set of 10 AGN templates, i.e. the limited AGN library, and a larger set of 100 random samplings from the 24,000-model AGN library, i.e. the full AGN library.  When the limited library is used, no random sampling is applied and the same 10 models are attempted in each run for the AGN component. Random sampling of the AGN library is only applied when using the full library, selecting 100 models to attempt with each fit.  The different settings of AGN models can be visualized in Figure~\ref{fig:torusmodels}, with their corresponding parameters listed in Table~\ref{table:agnparams}. To model the optical and IR emission components, from starlight and dust-reprocessed SF, respectively, we use 100 random samplings from each library for fitting.  In this section, we analyze the effects of AGN torus models on best-fit (1) MIR AGN Fraction, (2) AGN bolometric luminosity, and  (3)  IR-derived SFRs. 
 
 

 \subsection{AGN Decomposition with MIR Spectral fitting vs. Broadband SED Fitting}
\label{subsec:mirfracs}
Our 95-source sample is selected to contain sources with MIR AGN fractions between 0\% and 100\%, where these values were previously computed in \cite{Kirkpatrick2012} using Spitzer/IRS Spectroscopy. In addition to showing how our SED3FIT MIR AGN fractions are affected by torus models, we establish a calibration between the two distinct decomposition methods by comparing our results with \cite{Kirkpatrick2012} spectroscopic values. 

\subsubsection{Spectral MIR Decomposition Method in Kirkpatrick et. al (2012)}
\label{subsubsec:spec_decomp}

\cite{Kirkpatrick2012} decompose MIR spectroscopy into AGN and SF components,  computing MIR (rest-frame 5-12$\mu$m) AGN fractions for each source.  Their spectra is fit with a three component model including a local starburst composite to fit SF \citep{Brandl2006}, a pure power-law for AGN, and applying an extinction curve \citep{Draine2003} to the AGN component. Applying an extinction curve can identify and model Silicate absorption features, allowing Featureless and Silicate AGN to be distinguished for strong AGN. The decomposition method is outlined in detail in \cite{Pope2008}. 

 Figure~\ref{fig:torusmodels} demonstrates that NIR and FIR AGN SED features can affect the total MIR SED shape significantly. While \cite{Kirkpatrick2012} model broadband FIR SEDs with a blackbody curve  to compute total IR luminosity, their study did not use a multi-wavelength assessment of the AGN contribution over the entire mid-to-far IR range. Thus, a calibration between these two decomposition methods can shed light on how variation in AGN SED shapes shortward and longward of MIR spectral coverage may weight results from this MIR spectroscopic analysis.



\subsubsection{Mid-IR AGN Fraction Calibration}

 For all 95 sources in our sample, we compute our SED3FIT MIR AGN fractions by integrating the best-fit AGN template under the rest-frame 5-12$\mu$m range and dividing by the total SED, also in this range. While SED3FIT returns PDFs with each best-fit parameter,  we achieve a statistical distribution of best-fit quantities for all sources in this work by repeating each source's fit 100 times.  By repeating fits we create a set of solutions with statistical weights that test a generous variety of AGN models, building a distribution that is minimally biased to template assumptions. The corresponding best-fits thus represent a realistic range of solutions that are consistent with the observed photometry, shedding light on the true uncertainties of this approach. We continue to work with the entire family of solutions throughout the paper to detail the shape of these distributions and identify potential sources of bias or degeneracy. Thus, for each source we produce 100 repeated fits using the limited AGN library and 100 repeated fits using random sampling of the larger AGN library.

\begin{figure*}

\includegraphics[width=\textwidth]{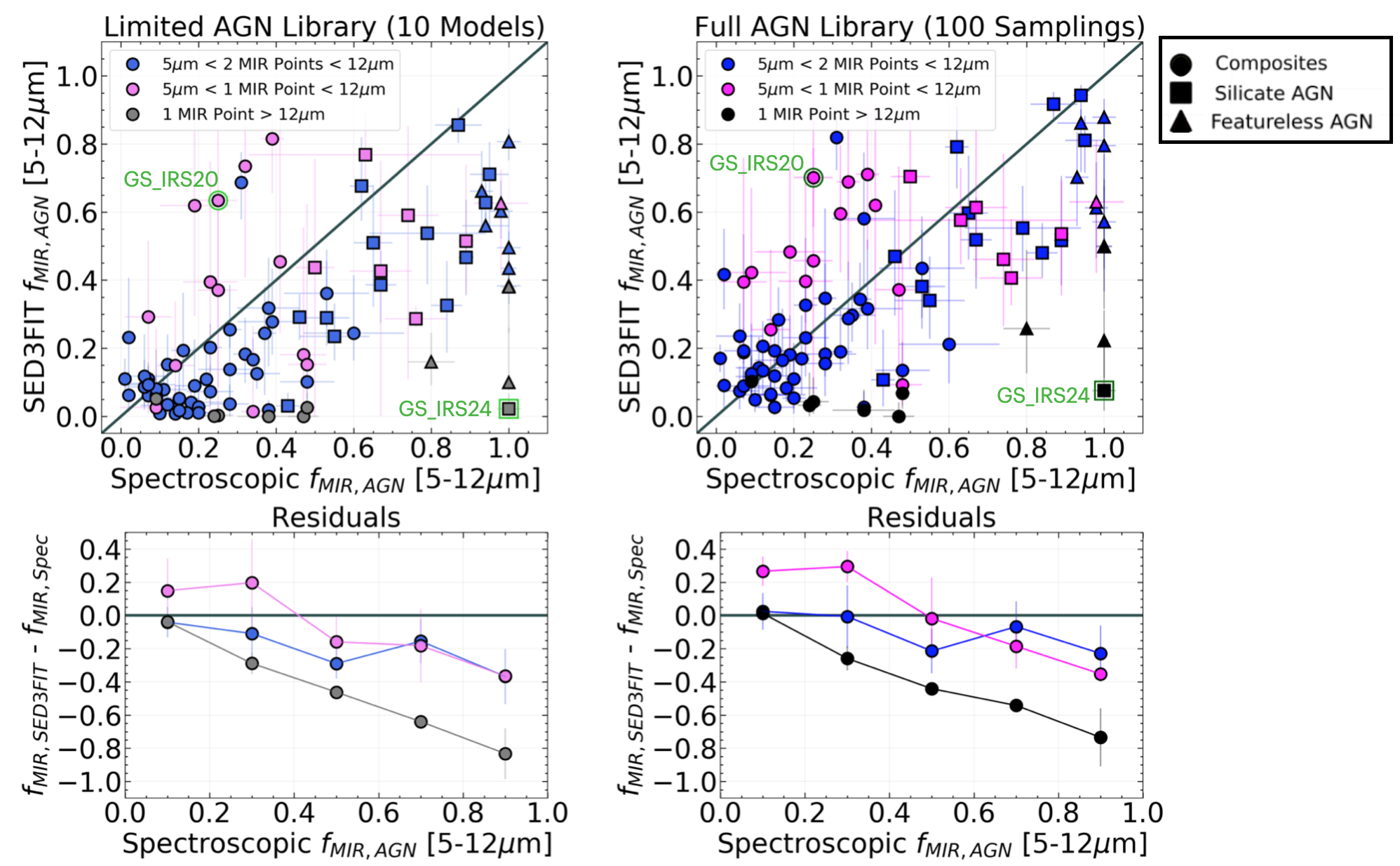}

\caption[2]{Comparison of rest-frame 5-12$\mu$m Mid-IR AGN fractions between our own SED decomposition with SED3FIT and the \cite{Kirkpatrick2012} study, using only mid-IR spectroscopy to derive AGN contribution. This comparison is shown for all 95 sources each ran with 2 sets AGN model choice in SED3FIT: (1) The limited set of 10 AGN templates, and (2)  100 random samplings from the full 21000-model AGN torus library. Focusing on the 5-12$\mu$m region, sources are classified by the amount of rest-frame constraints in this region. Sources in blue have both 16$\mu$m and 24$\mu$m detections within rest-frame 5-12$\mu$m. Sources in pink and black have only a 24$\mu$m detection; however objects in pink have this 24$\mu$m point inside the 5-12$\mu$m range and objects in black have this 24$\mu$m point redshifted longwards of 12$\mu$m. Circles represent composite AGN/SF galaxies, triangles are Featureless AGN, and squares are Silicate AGN, as spectroscopically classified in \cite{Kirkpatrick2012}.  Below we show the mean offset from the spectroscopic value as a function of MIR AGN fraction bin for each model setting. There are outlying two sources outlined in green; their example SED fits are shown in Figure~\ref{fig:outliers}}
\label{fig:bothcompare}
\end{figure*}

\begin{figure}

\includegraphics[width=.5\textwidth]{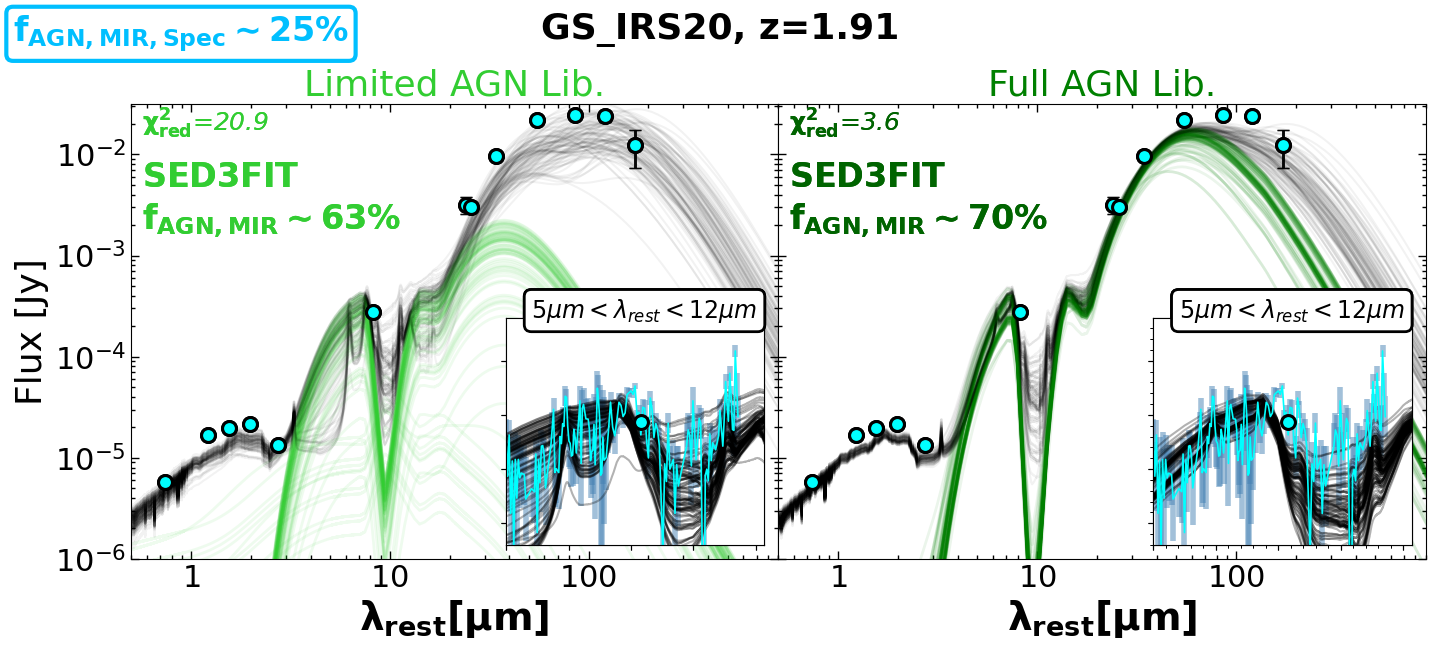}
\includegraphics[width=.5\textwidth]{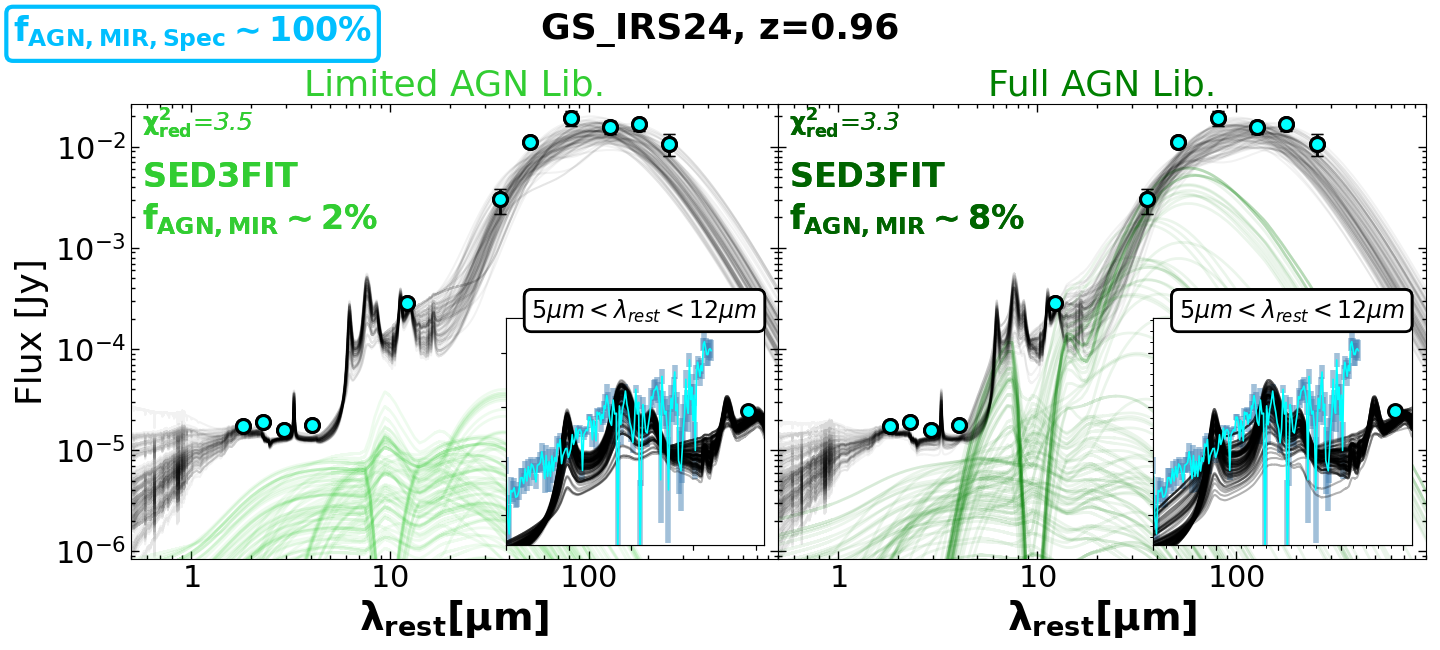}

\caption[2]{Two example fits with both the limited and full AGN library for outlying sources in the MIR AGN fraction comparison in Figure~\ref{fig:bothcompare}.  The top panel shows composite source GN\_IRS20, where SED3FIT computes a MIR AGN fraction >2x higher than the spectroscopically computed value of $\sim$ 25\% in \cite{Kirkpatrick2012}. This overestimation of AGN contribution is likely attributable to its single MIR constraint and its close proximity to the Silicate 9.7$\mu$m absorption feature. The bottom panel shows Featureless AGN GS\_IRS24, with a spectroscopically computed MIR AGN fraction of 100\%. The lack of broadband constraints in this region of the spectrum causes even this strong AGN to have a severely underestimated AGN contribution computed by SED3FIT of  < 10\%. Both fit examples show an inset plot of the Spitzer/IRS spectroscopy in cyan plotted over the MIR total SED best-fit models.}
\label{fig:outliers}
\end{figure}

 In Figure~\ref{fig:bothcompare} we show the full comparison between SED3FIT MIR AGN fraction values and the spectroscopic values for results using both the limited and full AGN library.  Values shown are the median and 1-sigma uncertainties for the distribution of 100 repeated best-fit values.  To illustrate the role of MIR sampling within this region, we differentiate sources based on the presence, absence, and rest-frame location of  16$\mu$m and 24$\mu$m data in reference to the 5-12$\mu$m range.  Sources that have only a 24$\mu$m constraint either have this data point redshifted within the  5-12$\mu$m range (pink points) or redshifted outside of this range (black points). In the latter scenario, objects have no constraints within this range and correspond to $z<1$. Sources outlined in green identify example outliers from the one-to-one relation that are discussed further in Section~\ref{subsubsec:outliers}, with corresponding fits also shown in Figure~\ref{fig:outliers}.

 Below each one-to-one comparison, we show the mean residual offset in AGN fraction for each scenario of MIR data location; the offset in comparison between the limited versus full AGN library is not statistically significant. However, the one-to-one plot indicates that for classified AGN, the limited library generally returns slightly lower MIR AGN fractions. 
 
 For the 31 Silicate and Featureless AGN with at least one MIR constraint (blue and pink points), our SED decomposition yields AGN MIR fractions lower by $\sim$ 25\% compared with the spectroscopic decomposition.  Featureless or Silicate AGN did not perform differently in the comparison; however, many featureless AGN had AGN MIR fractions of 100\% from \cite{Kirkpatrick2012}. These sources are more likely to be slightly underestimated by SED3FIT, which rarely returned MIR AGN fractions $>$ 95\%.   Strong AGN appear to be affected by MIR SED sampling only when the single 24$\mu$m point is redshfited $>$ 12$\mu$m (black points). In this case, SED3FIT systematically underestimates the AGN fraction by $\sim$ 40-60\% in both runs. 
 
 The general underestimation in MIR AGN fractions by SED3FIT could be explained by the averaged M82 starburst templates  and  power-law AGN model adopted in \cite{Kirkpatrick2012}. Their simplifiied treatment of starburst and AGN may have resulted in an underestimated MIR galaxy contribution and larger MIR AGN contribution. In SED3FIT, applying a complex set of AGN SED shapes introduces a wider variety of solutions more likely to result in lower MIR AGN fractions. 

 

 For composites with at least 2 MIR constraints in this region, the SED3FIT MIR AGN fraction approximately agrees with the spectroscopic values within uncertainties. When fitting is expanded from the limited to the full AGN library, there are  more composite outliers above the one-to-one line. These outliers consistently have a single constraint in this MIR range and overestimate MIR AGN fraction by  $\sim$ 25\%.  When no constraints are available within rest-frame 5-12$\mu$m, composites generally have lower MIR AGN fractions that follow the systematic under-estimation of MIR-strong AGN. \cite{Henry2019} carry out a similar analysis, comparing AGN contribution with spectroscopically-computed values using CIGALE to perform SED fitting. Their results support that mid-IR data is crucial to tease out AGN contribution, finding that z$\sim$ 3 objects have AGN contributions systematically lower as a consequence of the redshifted locations of their 24$\mu$m and 70$\mu$m data.

\subsubsection{Outliers explained: Systematic role of MIR SED sampling}
\label{subsubsec:outliers}

\begin{figure}
\includegraphics[width=\linewidth]{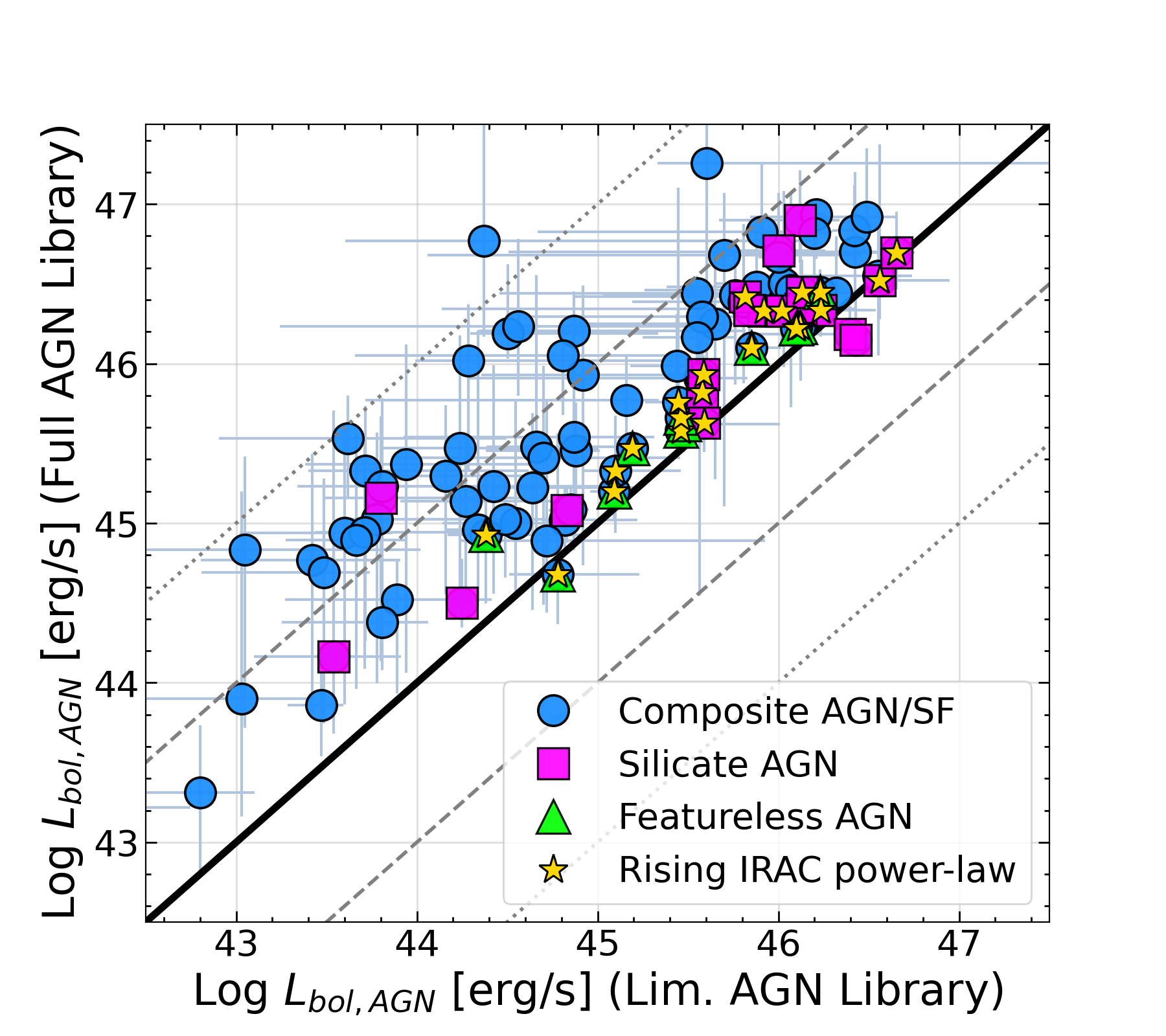}

\caption{Comparison of best-fit AGN Bolometric luminosity values for two different allowed AGN model settings for all 95 sources in our sample, 100 random samplings of the full AGN library (y-axis) and the limited set of 10 AGN models (x-axis).  Silicate AGN are shown as magenta squares, Featureless AGN are shown as green triangles, rising IRAC sources have a yellow star, and composites are shown in two shades of blue.  }
\label{fig:lbol_compare}
\end{figure}

It is apparent from Figure~\ref{fig:bothcompare} that certain gaps in MIR SED coverage are driving systematic differences in the MIR AGN fraction calibration.  Further, the systematics behave similarly in runs with the limited and full AGN library. In Figure~\ref{fig:outliers} we show 2 example outlier fits representing  (1) An outlying composite source with a higher SED3FIT MIR AGN fraction above the one-to-one line and (2) An outlying AGN with a significantly lower SED3FIT MIR AGN fraction than the spectroscopic value. 

Fits to composite source GS\_IRS20 in Figure~\ref{fig:outliers} reveal that high SED3FIT MIR AGN fractions are likely caused by the proximity of the 24$\mu$m constraint to the 9.7$\mu$m Silicate feature. In this case, the wide range of unsampled MIR SED results in the dominance of best-fit AGN models with deep Si absorption features. Not only do these models dominate, but this data gap causes the fits to have extremely high normalizations, yielding overestimated MIR AGN fractions. When a 16$\mu$m constraint is available, its presence likely prevents the fit from dominating this empty region, allowing for MIR AGN fractions more consistent with spectroscopic values. 

The example fits to Silicate AGN GS\_IRS24 in Figure~\ref{fig:outliers} show how the absence of rest-frame 5-12$\mu$m data render SED3FIT unable to constrain an AGN model with a high enough normalization to produce a reasonable MIR AGN fraction. The resultant set of best-fit AGN models are less constrained overall with low MIR AGN fractions that show a total MIR SED dominated by PAH features. 

Our analyses indicate that, when a Type-2 AGN model cannot be ruled out and is considered in fitting, objects with this MIR data setup (no rest-frame 5-12$\mu$m data) are difficult to disentangle with confidence. To test this systmatic effect, we run SED3FIT on the 65 sources that have both 16 and 24 $\mu$m data twice: With and without the 16$\mu$m constraint. We further elaborate on this analysis in the appendix but summarize results here. We find that the resultant MIR AGN fraction and $L_{Bol}$ are only significantly affected for composites and when the 16$\mu$m constraint is added to the rest-frame $\sim$ 5-6$\mu$m region.  Meaning, SED fitting results are stable to inconsistencies in MIR sampling so long as they have at least one rest-frame $\sim$ 5-6$\mu$m constraint and one longer wavelength rest-frame $\sim$ 12-18$\mu$m constraint. When we consider this variety of AGN models with Si absorption,  the fitting of composite AGN/SF sources that lack a rest-frame $\sim$ 5-6$\mu$m constraint, is likely to systematically over-estimate MIR AGN fraction by $\sim$ 25\% and overestimate $L_{Bol}$ by 1-2 dex. This leaves sources at  $z \sim 1.5-2$ most susceptible to systematic effects.

\subsection{Effect of torus models on AGN Bolometric Luminosity }
\label{subsec:lbol}


We next explore the effects of AGN torus models on AGN Bolometric luminosity, $L_{Bol}$, where $L_{Bol}$ values are  returned as a best-fit parameter in SED3FIT. In Figure~\ref{fig:lbol_compare} we show the comparison of best-fit  $L_{Bol}$ between runs with a limited and full AGN torus library. We also show the residual difference in $L_{Bol}$ as a function of redshift in the right panel.  It is important to note that a portion of $z \sim 2$ composite sources with a single 24$\mu$m constraint are subject to systematic fits for both runs. This is due to the proximity of this data point to the 9.7$\mu$m Silicate absorption line, as discussed in the previous section.  Since both sets of AGN model fits see a similar effect, the ramifications of this systematic on additional best-fit parameters are beyond the scope of this work. For the remainder of this analysis, we make the assumption that disparities in results are  purely attributable to the two different sets of AGN models tested in our in UV-FIR SED fitting. 

Figure~\ref{fig:lbol_compare} shows that strong AGN ($f_{MIR,AGN} > 50\%$)  see the least residual difference in best-fit $L_{Bol}$ between the limited and full AGN model choices. The correlation around the one-to-one line is also tighter for rising IRAC sources, where the full library returns $L_{Bol}$ values greater by $\sim$ 0.5 dex or less compared with the limited library. The residual offset is more severe for weaker AGN, namely the composite sources ($f_{MIR,AGN} < 50 \%$) at z$\sim$ 1. In these sources, the full AGN library can result in higher $L_{Bol}$ values  by as much as $\sim$ 2 dex compared with the limited library.  This difference gradually decreases as redshift approaches 0.5, though it is unclear whether this is a systematic effect of MIR SED sampling.


These findings have significant implications on the robustness of SED decomposition for samples with similar broadband data availability.  Decomposition results become significantly more sensitive to permitted AGN models when the IRAC power-law dips below monotontically rising.  While sources with rising IRAC power-laws are commonly in MIR color selection wedges, even many color-selected AGN do not exhibit this power-law feature. Once we also consider the plethora of composite  sources in IR-samples, this leaves a significant portion of sources subject to either large uncertainties or systematic effects on AGN luminosity.  We explore the effects of this on AGN incidence rates and bolometric luminosity functions in the discussion.

\subsection{AGN contamination to the FIR SED: How do torus models affect SFR?}

\begin{figure*}
\includegraphics[width=.53\textwidth]{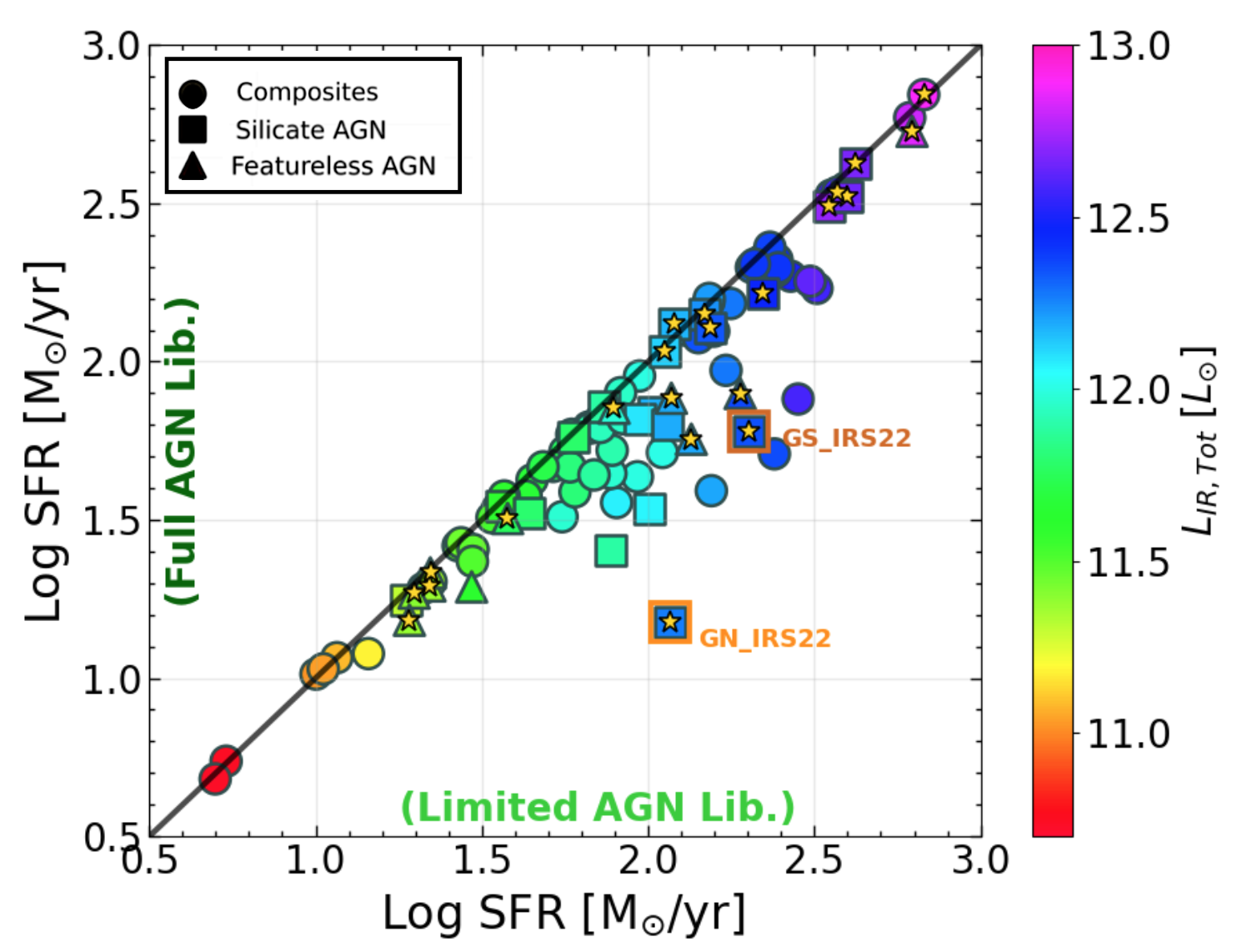}
\includegraphics[width=.44\textwidth]{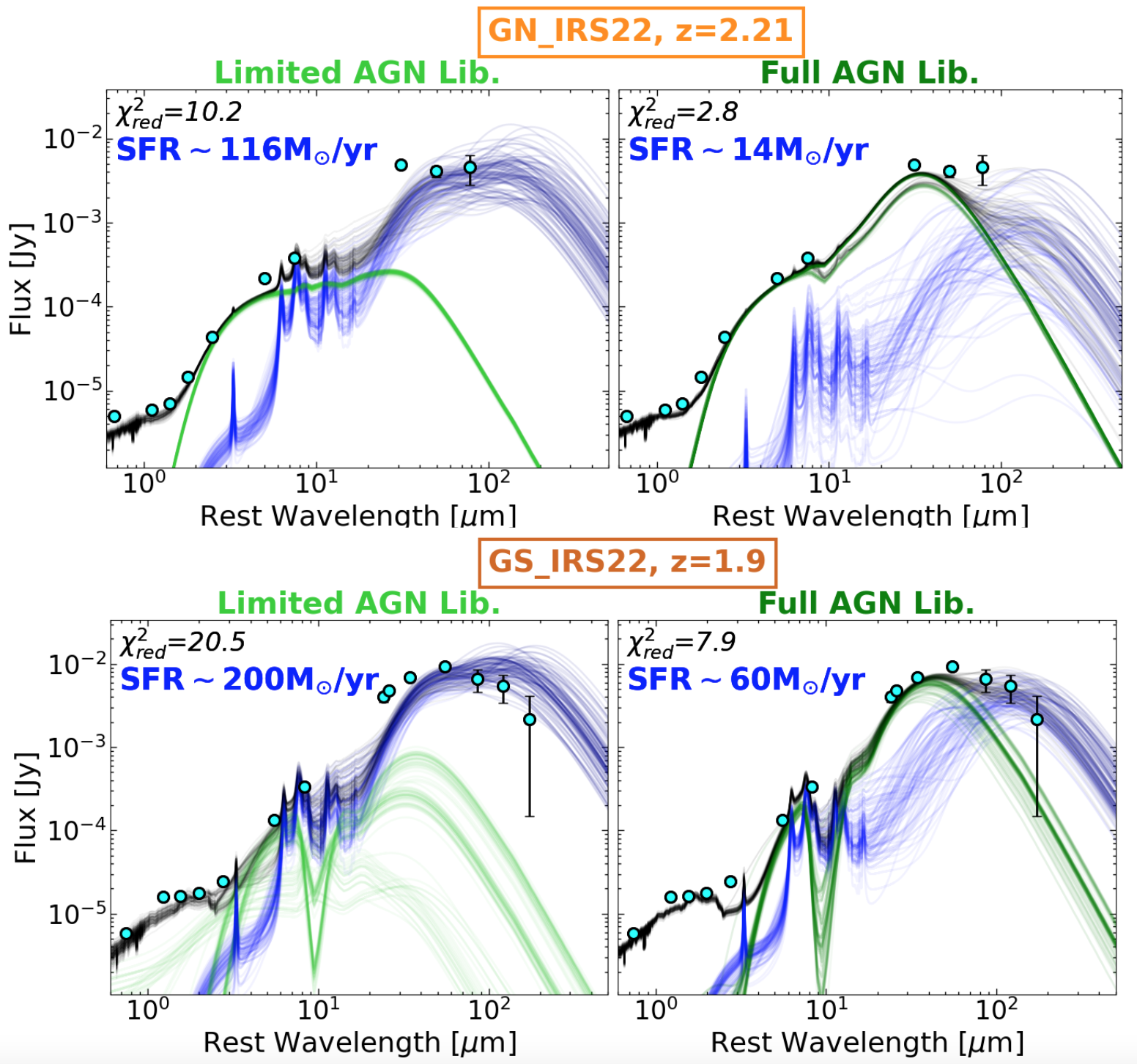}
\caption{SFR comparison for SED3FIT results between the limited and full AGN library. Stars indicate sources with a rising IRAC power-law and the colorbar shows the total $L_{IR}$ for each source. Two AGN with SFRs deviating from the one-to-one line are identified, GS\_IRS22 and GN\_IRS22, with their corresponding SED fits shown beside for each AGN library run. The examples illustrate that fits returned by the full AGN library have improved fit quality, i.e. lower $\chi_{red}^{2}$, and higher AGN contributions to the FIR SED than the limited library, driving the difference in SFR.   }
\label{fig:sfrs}
\end{figure*}

 
 The FIR regime is an indirect tracer of star formation under the assumption that the FIR emission originates purely from stellar UV radiation, without AGN contamination. This assumption rests on temperature differences, where the dust temperature associated with stellar emission is cooler than warm AGN-heated dust and peaks at longer FIR wavelengths \citep{DaleHelou2002,Harrison2012}. 
 
 Recently, however,  the extent of FIR emission attributable to AGN has been brought into question, and many are revisiting the accuracy of SFR estimates that disregard AGN contribution \citep{Symeonidis2016, Roebuck2016, McKinney2021,Symeonidis20201}. Not accounting for AGN contribution may result in vastly over-estimated SFRs for AGN-dominated galaxies with the highest IR luminosities $> 10^{13}$ L$_{\odot}$. Though the majority of IR-selected sources are not dominated by AGN  ($\sim $25\% incidence rate), when composite sources are also considered, an additional $\sim$ 50\% of sources may contain FIR AGN contamination \citep{Kirkpatrick2017}. 

We consistently account for AGN contribution, but explore whether IR-derived SFRs for our sample are modified significantly by the different AGN torus models adopted in fitting. In the left panel of Figure~\ref{fig:sfrs} we show the difference in SFR  between runs with the limited and full AGN torus library. Sources with a rising IRAC power law are indicated and the colorbar represents total IR luminosity.  We find that increasing our choice of AGN models to the highest optical depths can reduce the SFR by a factor of 2-5x in $\sim$ 10\% of sources, where these sources have $L_{IR} > 10^{12}$ $L_{\odot}$. For one AGN source, fitting with the full AGN library reduces the SFR by a factor of 8x.


In the right panel of Figure~\ref{fig:sfrs} we show two example fits of rising IRAC AGN with SFRs strongly affected by the AGN model parameter space. We show both the 100 repeated best-fit AGN templates and  SF templates for each run to illustrate how the two components affect one another in this three-component fitting. For both examples, the extended library produces improved fit quality i.e. lower reduced $\chi^{2}$ values.  Visually, we also see that the broadband points around rest-frame 20-30$\mu$m are better-fit overall when the AGN template fills in this warm-dust region of the total SED. As the preferred AGN template shifts to a rest-frame 20-30$\mu$m peak with greater intensity, the dust-reprocessed SF contribution diminishes to yield a lower derived SFR. We also note that all 95 sources saw improved fit quality with lower reduced $\chi^{2}$ values when fit with the full AGN library. 

Our results demonstrate that adopting a simplistic range of AGN SED models can  overestimate SFRs if alternative AGN models are not explored for improved fit quality. Since this discrepancy is not widespread, it is not likely to affect SF luminosity functions (LFs) or propagate to studies of star formation rate density (SFRD) evolution in large samples.

 \section{Discussion}
 \label{sec:discussion}

\subsection{AGN incidence rate in IR-samples and $L_{Bol}$ Luminosity Functions}
\label{subsec:agn_incidence}



Characterizing the AGN bolometric luminosity of objects with co-existing AGN and SF activity is crucial for obtaining a complete census of supermassive black hole accretion history \citep{Lacy2020}. To target this composite AGN/SF galaxy population, our 95-source sample spans a dynamic range in AGN strength, with spectroscopic MIR AGN fractions between 0\% and 100\%.  In Section~\ref{subsec:lbol} we show that, for composite sources, decomposing SEDs with a limited AGN library can return $L_{Bol}$ values lower by $\sim$ 1-2 dex compared with results using the full AGN library.

If discrepancies in $L_{Bol}$ are widely propagated throughout a sample, varied results may determine which sources are included or omitted from an AGN-focused study, affecting the AGN incidence rate. The AGN incidence quantifies the fraction of a galaxy sample that hosts an AGN component; however, the method used to compute AGN contributions within individual galaxies will affect this quantity. Recently, \cite{Symeonidis20201} address discrepancies in literature estimates of the AGN incidence fraction within IR-selected samples.  \cite{Symeonidis2014} find a $\sim$ 20\% AGN incidence for a sample of IR-selected galaxies using quantities such as hardness ratio, X-ray variability, optical excitation lines, and MIR colour to determine AGN presence. \cite{Delvecchio2014} use broadband AGN SED decomposition to estimate a larger AGN incidence of 37\%  for a 160$\mu$m-selected sample. \cite{Symeonidis20201} credit the higher incidence rate in \cite{Delvecchio2014} to their broadband decomposition method, asserting that SED fitting is more sensitive to composite sources than the calculations employed in \cite{Symeonidis2014}.  
 
These incidence fractions can be further broken down into relative contributions from MIR-strong AGN and composites. \cite{Gruppioni2013}  use a 160$\mu$m-selection with SED fiting and find that AGN-SB composites comprise $\sim$50\% of the population while AGN-dominated sources comprise 10\%. In this work, the AGN + composites comprise $\sim$60\% of our  24$\mu$m-selected sample (95 AGN + composite  sources studied in this paper out of the full 151-source sample described in Section~\ref{sec:sample}). \cite{Sajina2012} find a similar estimate in their 24$\mu$m-selected sample, stating that as much as 70\% of the population can contain AGN contribution \citep{Kirkpatrick2017}.

 \begin{figure}
\includegraphics[width=.5\textwidth]{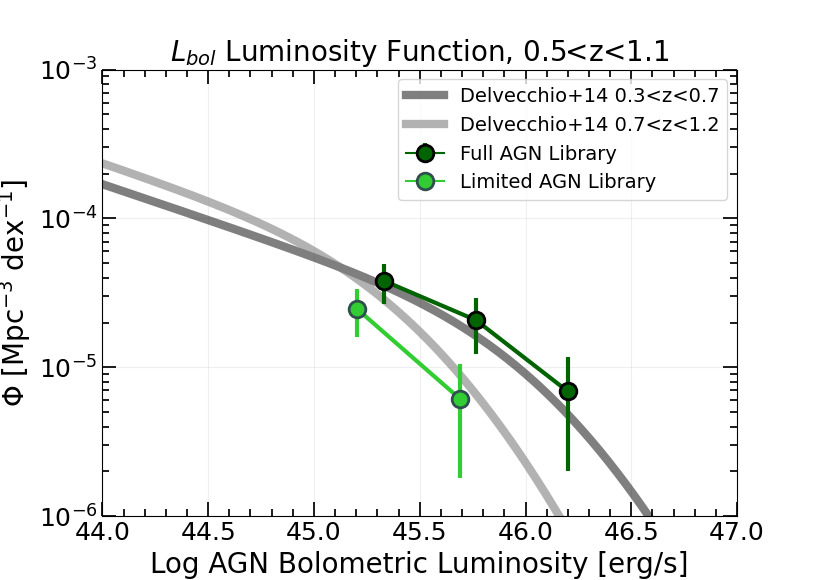}
\includegraphics[width=.5\textwidth]{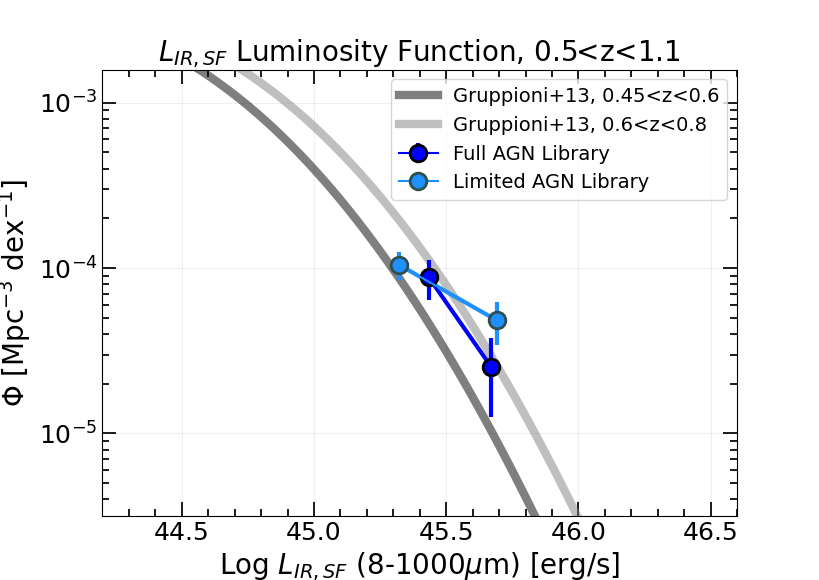}
\caption[2]{\textit{Top:} AGN bolometric luminosity function (LF) in the $0.5 < z < 1.1$ range for our sample using results from two different allowed AGN model settings: (1) 100 random samplings of the full 21000-model library, shown in dark green, and (2) a restricted set of 10 templates where no random sampling is applied, shown in light green.  For comparison, we show two luminosity functions modeled by the Shechter function from \cite{Delvecchio2014} that overlap in our chosen redshift range, where our wide redshift bin of $0.5 < z < 1.1$ is chosen due to low number statistics. \textit{Bottom:} The $L_{IR,SF}$ luminosity functions including the star forming galaxies with both AGN model setting. Results from fitting with the full AGN library are shown in dark blue, with results from the limited library of 10 models shown in light blue. These LFs are shown with two $L_{IR,SF}$ Schechter function trends from \cite{Gruppioni2013} within our redshift range. Both LF demonstrations are meant to show the differences purely driven by adopted AGN templates, as opposed to a robust luminosity function estimate, since this sample does not have a straightforward selection function.}
\label{fig:lumfunction}
\end{figure}

 \begin{figure*}
\centering
\includegraphics[width=\textwidth]{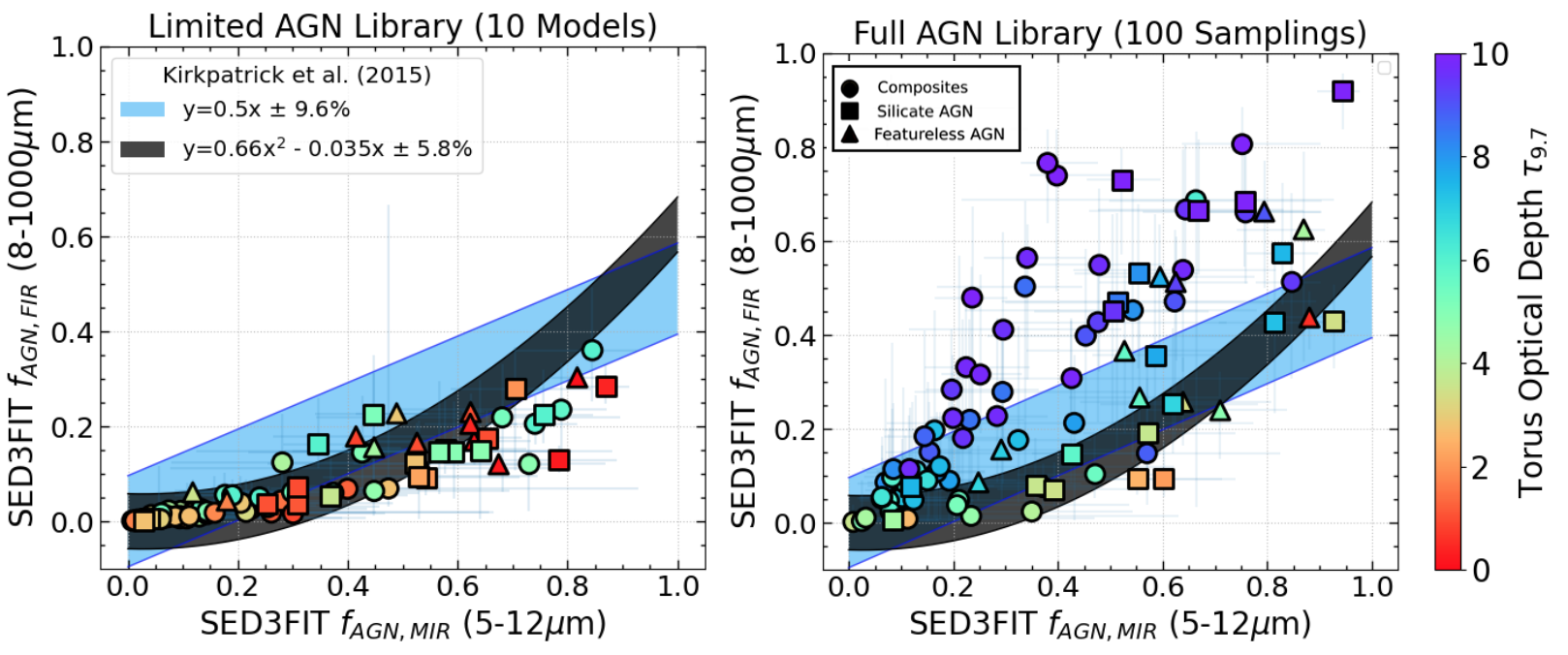}
\includegraphics[width=.48\textwidth]{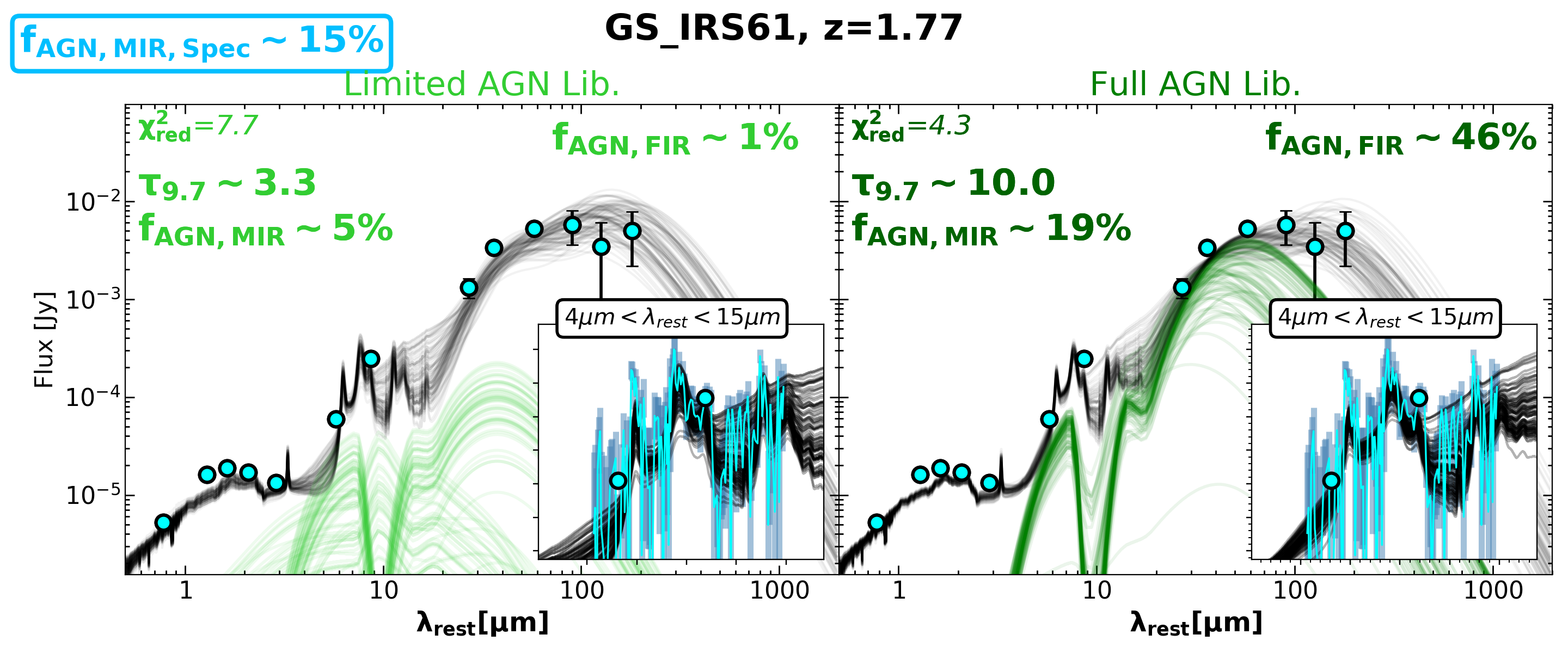}
\includegraphics[width=.48\textwidth]{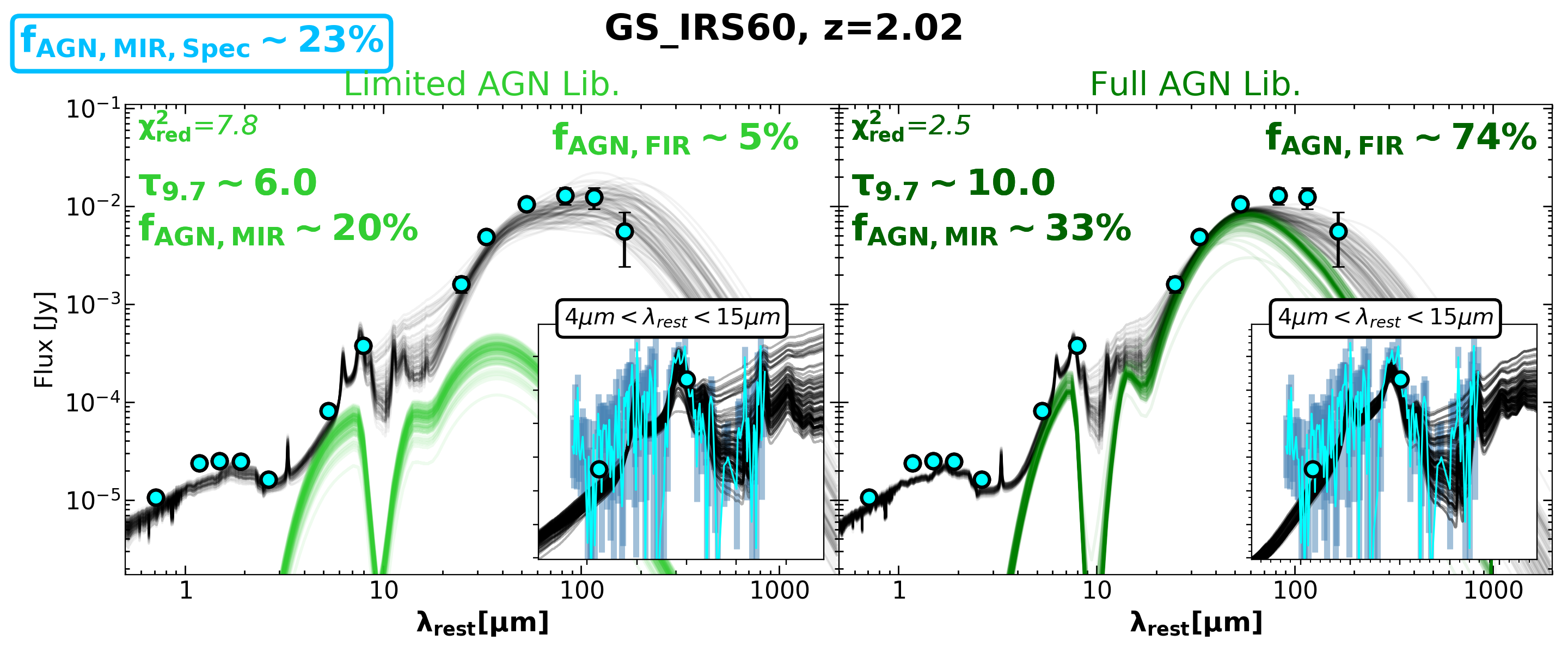}
\includegraphics[width=.48\textwidth]{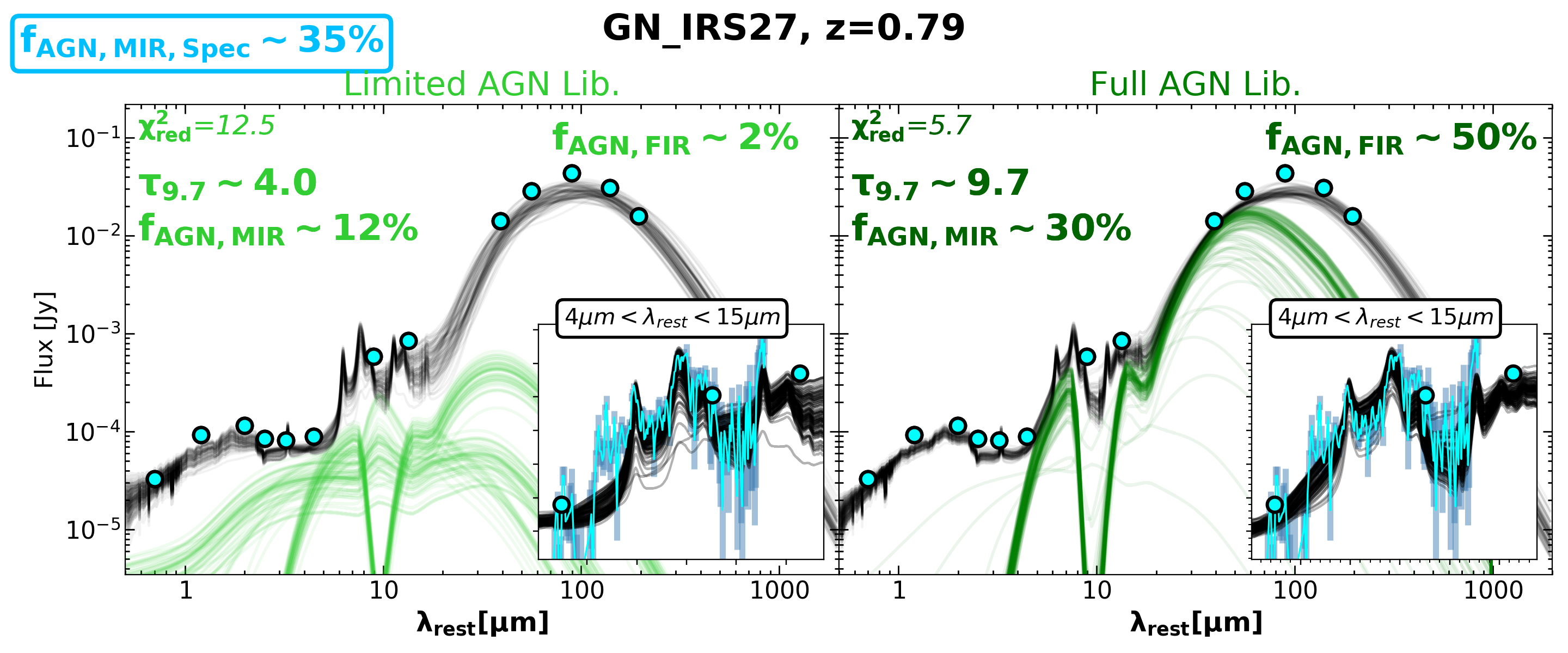}
\includegraphics[width=.48\textwidth]{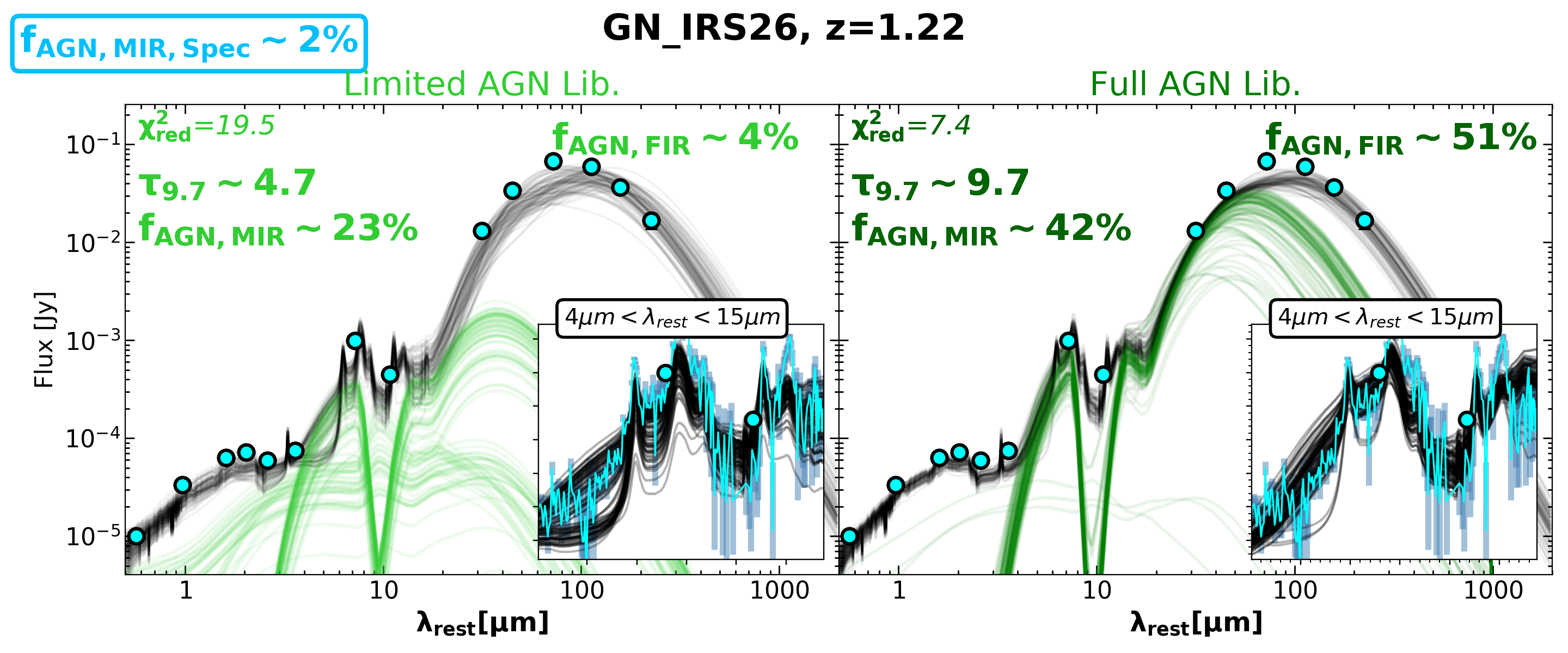}
\caption[1]{\textit{Top panels:} The relationship between AGN MIR (rest-frame 5-12$\mu$m) and AGN FIR (rest-frame 8-1000$\mu$m) fractional contribution, where the MIR AGN fraction and FIR AGN fraction is derived in this work using SED3FIT. FIR AGN fractions are shown for both sets of runs using the limited set of 10 AGN templates (left) and 100 random samplings of the entire AGN library (right).  Sources are color coded by their best-fit torus optical depth, $\tau_{9.7}$, where the maximum allowed $\tau_{9.7}= 10.0$  for the full library and 6.0 for the limited library. Markers are shaped as circles, squares, and triangles representing composites, Silicate AGN, and Featureless AGN, respectively. The dashed and shaded blue and black lines represent the empirically estimated linear and quadratic  MIR-FIR AGN fraction relations in \cite{Kirkpatrick2015}, respectively. \textit{Bottom panels:} Four example fits to sources shown with Spitzer MIR spectroscopy overplotted. }
\label{fig:mirfir}
\end{figure*}

The AGN incidence fraction, in tandem with conflicting estimates of $L_{Bol}$ (see Figure~\ref{fig:lbol_compare}), may also affect the shape of $L_{Bol}$ luminosity functions (LFs). If the number of sources included in the sample increase with the inclusion of more composites, the propagated effects on $L_{Bol}$ LFs will depend on the added $L_{Bol}$ values. For example, the addition of low-luminosity AGN to a sample may not affect the LF  significantly.  Although these objects may have previously been hypothetically omitted, they are likely to contribute to the incomplete portion of the LF that cannot be integrated to derive black hole accretion rate. However, if composite sources are more bolometrically significant than previously assumed, this offset could translate to the bright end of the AGN bolometric luminosity function more significantly,  affecting the integral quantity used to derive BHARD evolution with redshift. It is important to note that since SFR is also sensitive to the applied AGN templates, the $L_{IR,SF}$ luminosity functions and the SFRD would also be affected. Although, given that strong AGN are most affected by SFR differences (see Figure~\ref{fig:sfrs}), the low number statistics of these objects means that the $L_{IR,SF}$ LFs should not be affected as significantly by AGN template as the $L_{Bol}$ LFs. 

\cite{Delvecchio2014} compute the only IR-derived BHARD to date with 160$\mu$m-selected sources in the GOODs and COSMOS fields. Their 37\% AGN incidence, including 1100 AGN total, is estimated by comparing $\chi_{red}^{2}$ values between fits with and without an AGN component included to assess how frequently the AGN component improved the broadband SED fit. Their study uses the same theoretical AGN templates \citep{Feltre2012,Fritz2006} as this paper and includes a combination of sources with MIR detections at both 16$\mu$m and 24$\mu$m as well as only 24$\mu$m.  A key difference in their work is the omission of $\tau_{9.7} >$ 6.0 AGN models from fitting, analogous to the optical depth cutoff in our limited AGN library.

We have shown that allowing higher optical depth AGN models can result in AGN luminosities greater by nearly 2 orders of magnitude with improved fit quality for composite sources. Thus, if not much is known about these sources a-priori besides MIR color, allowing higher optical depth models in SED fitting may introduce higher $L_{Bol}$ solutions.  This  would (1)  increase the AGN incidence rate from \cite{Delvecchio2014} and (2) potentially modify the resultant AGN bolometric luminosity functions and BHARD. Moreover, since this effect is most prominent for the abundance of composite sources outside of typical color selection wedges, its propagation may be widespread.


We express the resultant differences in best-fit AGN luminosity as two separate AGN luminosity functions, from both the full and limited set of AGN templates, for  $0.5 < z < 1.1$  sources. We do the same for $L_{IR,SF}$ luminosity functions, where both are constructed using the $\frac{1}{V_{max}}$ method. However, since our sample does not apply straightforward selection function, we emphasize that these luminosity functions are meant to be purely illustrative of differences driven by allowed AGN models, rather than a robust final calculation. Objects included in this study require $S_{24}>$ 100 $\mu$Jy, as opposed to the respective survey 24$\mu$m flux limits (5$\sigma$ $S_{24,lim}$ =70$\mu$Jy for ECDFs and 5$\sigma$ $S_{24,lim}$ = 30$\mu$Jy for GOODS-N). As a result, there are additional objects in these fields below $S_{24,lim}$ that are not accounted for. Sources also require Spitzer/IRS spectroscopic observations; combining both GOODS-N and ECDFS fields, our IRS sample includes $\sim$ 15\% of all sources in these fields above the 100$\mu$Jy  flux threshold. The derived luminosity functions are scaled up accordingly to account for these missed objects.

In Figure~\ref{fig:lumfunction} we show the $L_{Bol}$ and $L_{IR,SF}$ luminosity function estimates for sources with $0.5 < z < 1.1$. For reference we also show the corresponding LFs in \cite{Delvecchio2014} and \cite{Gruppioni2013} in two overlapping redshift ranges. Our $L_{Bol}$ LFs are roughly complete to Log $L_{Bol} \sim 45$ erg/s and the apparent difference in bright-end slopes demonstrates that increasing the parameter space of the applied AGN models can significantly impact the AGN LFs. However, we caution that all LFs constructed here are estimates with corrections applied based on our non-trivial sample selection. Their purpose is to demonstrate how different AGN model choices can affect the LF bright end significantly. The $L_{IR,SF}$ luminosity functions shown in the bottom panel of Figure~\ref{fig:lumfunction} are constructed by also adding in SED3FIT results for the 49 the star-forming galaxies in the parent sample. Although analysis of these 49 SFGs is beyond the scope of this paper, when fit with SED3FIT we find these sources have a mean $f_{MIR,AGN} \sim 3\%$ with the limited AGN library and a mean $f_{MIR,AGN} \sim 10\%$ with the full AGN library. The LFs show a slight offset driven by AGN model, where the full AGN library values show a steeper bright-end slope. When applying this method to larger samples, this offset will become less statistically significant as the contribution from strong AGN in the redshift bin decreases and the fraction of SFGs with robust $L_{IR,SF}$ estimates increases. 

It is clear that composite AGN-SB sources are statistically significant in both IR and MIR selected samples. Without accounting for these objects it is difficult to confidently assess the black hole accretion history of the universe in conjunction with the simultaneous stellar mass buildup. Though there are many possible solutions to these fits caused by sparse SED sampling, we cannot rule out the possibility that composite sources may be much more bolometrically significant when a larger AGN library is applied to fitting. As a result, their inclusion and SED decomposition method may affect IR-selected AGN luminosity functions substantially.

\subsection{$f_{AGN,FIR}$ vs. $f_{AGN,MIR}$ Relation }


It is well-established that excess MIR emission from an AGN component originates from a warm circumnuclear dusty torus; however, for sources containing an ambiguous mixture of SF and AGN activity, the source of galactic FIR emission is more puzzling. Observed FIR emission has been absorbed and reprocessed into the FIR by dust from shorter wavelengths. This makes it difficult to identify the true origin of  FIR photons, since the information is effectively erased during the reprocessing of emission. In this section, we explore the relationship between MIR AGN fraction and FIR AGN fraction to assess the effects of including higher optical depth AGN models in fitting.  Best-fit torus optical depth, $\tau_{9.7}$, sheds light on the extent of circumnuclear dust obscuration in our sample, revealing objects that may be candidates for extremely luminous buried AGN.

\subsubsection{Comparison with two-temperature FIR decomposition in Kirkpatrick+15 }


We compare the relationship between MIR and FIR AGN fraction in our sample, where the AGN FIR fraction is derived directly from SED3FIT by integrating best-fit AGN and total templates under the rest-frame 8-1000$\mu$m range. While our results permit and test a wide range of AGN SED shapes and corresponding physical parameter space (see Table~\ref{table:agnparams}), \cite{Kirkpatrick2015} provide an  estimate of FIR AGN contribution empirically. They create an AGN-only template by applying a two-temperature blackbody fit to an AGN+host galaxy template. Under the assumption that the cold dust component of this two-temperature fit is entirely due to the host galaxy, subtracting its emission from the AGN template in theory produces an AGN-only template. They perform the decomposition by fitting simultaneously this AGN-only template and a $z\sim$1  SF SED. As a result, they derive two relationships, linear and quadratic, between AGN MIR and AGN FIR Fraction. It is important to note that the trends derived in \cite{Kirkpatrick2015} use an AGN template built from a sample consisting of predominately Type-1 AGNs. Compared with Type-2 AGNs, Type-1 AGN templates will have less MIR dust absorption and subsequently less reprocessed FIR AGN emission.

In Figure~\ref{fig:mirfir} we show the linear and quadratic relations derived in \cite{Kirkpatrick2015} for $f_{AGN,FIR}$ vs. $f_{AGN,MIR}$ along with our own SED3FIT values for each AGN model setting. Sources are colored by  best-fit torus optical depth, shown in the color bar, to illustrate that this parameter is most likely responsible for the differences between our results and these linear/quadratic relations. The full AGN library reaches a maximum $\tau_{9.7}$  = 10.0 while the limited library reaches a maximum $\tau_{9.7}$=6.0. Below, in Figure~\ref{fig:mirfir} we show four example fits for MIR-weak AGN that saw a much higher FIR AGN fraction with the full AGN library. These example fits also show a zoomed inset with the Spitzer spectroscopy plotted over the total best-fit MIR SEDs; it is important to point out that these SED3FIT results are largely consistent with the MIR spectroscopy.  Meaning, given all available data there is no reason to rule out the majority of these SED fiting solutions, for both limited and full AGN libraries, despite contrasting values. 

Focusing on the effects of varying torus models, a key takeaway from this figure is that in strong AGN ($f_{MIR,AGN} > 50\%$), the limited library returns significantly lower $f_{FIR,AGN}$ than the \cite{Kirkpatrick2015} relation. For the limited library, strong AGN have an average $f_{FIR,AGN}$ $\sim$ 15-20\% computed with SED3FIT, while the empirically derived relations predict $f_{FIR,AGN}$ $\sim$ 40-50\%. Further, no solutions from the limited AGN library reach the maximum predicted FIR AGN fraction of $\sim$ 60\%. This is compelling, as the empirical templates deriving this relation are similar in FIR emission shape to the limited library that is also widely adopted in the literature. For the full AGN library, about half of the strong AGN follow the empirical trend values while half return higher FIR AGN fractions by $\sim$ 20\%, still following a similar trend in shape. Overall the comparison highlights the important role played by both method and AGN template in SED decomposition, emphasizing that extrapolations based on such trends may be biased to the SED fitting approach. Our results suggest that it is unclear whether there even exists some predictable trend between  $f_{AGN,FIR}$ and $f_{AGN,MIR}$, due to the role of optical depth.

\subsubsection{Role of $\tau_{9.7}$: Do composite galaxies have more AGN FIR contribution than previously expected?}
\label{sec:roleoftau}

The most apparent difference between the empirical trend and our SED3FIT results are seen when $\tau_{9.7} \sim 10$ for the full AGN library. Over the full range of MIR AGN strengths, these high optical depth torus models yield FIR AGN fraction solutions that completely deviate from the predicted trend in \cite{Kirkpatrick2015}. Figure~\ref{fig:mirfir} shows example fits to composite sources that contribute to the high optical depth trend deviation. These examples, GS\_IRS61, GS\_IRS60, GN\_IRS27, and GN\_IRS26 also have SED3FIT $f_{MIR,AGN}$ results consistent with $f_{MIR,AGN,Spec}$, supporting the plausibility of such high optical depth solutions when we increase AGN model sampling. The four example fits in Figure~\ref{fig:mirfir} show lower reduced $\chi^{2}$ values by a factors of $\sim$2 when fit with high optical depths in the full AGN library and have adequate photometry near the peak of the AGN SED to justify the inclusion of a strong FIR AGN component.

In the four examples in Figure~\ref{fig:mirfir}, the full AGN library returns improved fit quality and SED3FIT $f_{FIR,AGN} \sim 50\%$ or higher. GS\_IRS61 and GN\_IRS26 represent weak composites with $f_{MIR,AGN} \sim 15\%$ and $\sim 2\%$, respectively. The \cite{Kirkpatrick2015} trend posits that these sources should have $f_{FIR,AGN} \sim 5\%$ while the presence of an optically thick torus would result in $f_{FIR,AGN} \sim 50\%$ via SED fitting. Similarly high $f_{FIR,AGN}$ values are found using the full AGN library in SED fiting for GS\_IRS60 and GS\_IRS27, moderately strong composites with $f_{MIR,AGN} \sim 23\%$ and $\sim 35\%$, respectively. The trends estimate these sources should have a much lower $f_{FIR,AGN} \sim 10\%$. These fits, along with the plotted sources that behave similarly, demonstrate the feasibility of maintaining a low MIR AGN fraction with a high FIR AGN fraction for heavily obscured, optically thick torus models. 

Many literature estimates of $f_{FIR,AGN}$ for composites are comparable to the low values predicted by the \cite{Kirkpatrick2015} trend.  Typically, AGN models used to compute these low $f_{FIR,AGN}$ $\sim$ 10\%-20\% estimates are also derived empirically, as in \cite{Kirkpatrick2015}. However, the theoretical AGN library of \cite{Fritz2006} used in this work is widely used in literature applications of SED fitting. \cite{Ramos2020} use CIGALE to decompose UV-FIR SEDs of nearby AGN, starbursts, and interacting galaxies. They adopt the \cite{Fritz2006} AGN library including $\tau_{9.7}=10.0$ models and find many best-fits similar to the full AGN library examples in Figure~\ref{fig:mirfir}. These solutions were predominately found in interacting galaxies but showed a best-fit AGN model with a deep Silicate absorption feature, yielding low MIR AGN contributions, and strong FIR emission with $f_{FIR,AGN}$ $\sim$ 30\%-50\%.

\cite{Roebuck2016} also find higher FIR AGN contributions for composites, exceeding the common literature estimates of $\sim 10\%-20\%$. The authors compute simulated AGN FIR fractions for SFGs, Composites, and MIR-strong AGN to compare with empirical values adopted from the two \cite{Kirkpatrick2015} trends. They find the largest deviation in composite galaxies, where simulations predict FIR AGN fractions $\sim$ 3x larger than empirical values. Additionally, the best-fit extinction parameters for these objects position them in the optically thick regime. 

\cite{Roebuck2016} conclude that the method in \cite{Kirkpatrick2015} may underestimate AGN contribution longward of 100$\mu$m and that bolometrically significant AGN sources may be mistaken for MIR-weak composites due to a heavily obscured MIR continuum that is reprocessed into the FIR. Their result is consistent with our findings, where our composites with best- fit $\tau_{9.7} \sim 10$ in particular have considerably higher FIR AGN fractions $\sim$ 2-5x larger than the predicted trend in \cite{Kirkpatrick2012}. 

Overall, conflicting FIR AGN contribution estimates point to the difficulty of uncovering potentially buried AGN in composite galaxies.  Specifically,  fitting SEDs with theoretical torus models that simulate an optically thick torus can yield results with stronger FIR AGN contribution. Sub-mm observations of ultra-luminous IR galaxies (ULIRGS) in \cite{Chen2020} show evidence of a compact obscuring material near the nucleus. Referred to as compact obscuring nuclei (CONs), if these objects are widespread in ULIRGs they support the notion that  an optically thick nuclear region may be common, and hidden in the reprocessed FIR SEDs of IR-bright composite galaxies.


~~~~~~~~~~~~~~~~~~~~~~~~~~~~~~~~~~~~~~~~~~~~~~~~~~~~~~~~
\subsection{X-ray Luminosity as an additional probe of obscuration }


 
 The obscured AGN phase of galaxy evolution proposed by \cite{Sanders1998} is thought to represent an epoch where SMBHs accrete the bulk of their mass behind thick screens of dust.  The diffuse X-ray background also suggests nearly 50\% of AGN are highly obscured; thus, understanding dusty optically thick AGN properties is necessary for studies of universal SMBH accretion history and universal mass assembly.
 
 In obscured AGN, large columns of gas and dust are concentrated in the nuclear region that can interact with black hole radiation mechanisms at various wavelengths: UV-FIR photons can be absorbed by obscuring dust and X-ray photons can be absorbed by obscuring gas.  For typical gas-to-dust ratios, an obscured, Compton-Thick AGN corresponds to a dust extinction screen of $A_{V} \sim 5-10$ mag and gas column density $N_{H} > 10^{22} \rm cm^{-2}$ \citep{Perna2018,Ricci2017a}. Since AGN obscuration correlates with X-ray photon absorption, sources under-luminous in X-ray emission, or undetected in X-ray, are likely to host high levels of nuclear obscuration.

\subsubsection{X-ray Data}

 Our goal is to investigate whether potentially obscured sources in our sample, best-fit with high torus optical depths, are under-luminous in X-ray for their IR luminosity compared with remaining optically thin sources. For the 95 sources in our sample, we glean X-ray data by positionally matching sources using a 1.5`` radius to the Chandra catalogues for GOODS-N \citep{Luo2017} and ECDF-S \citep{Xue2016}.   For objects with X-ray matches, we extract the intrinsic 0.2-8 keV X-ray luminosities from these publicly available catalogues and convert this luminosity to the 2-10 keV X-ray luminosity, assuming a photon index $\Gamma$ =1.9. 55 of our 95-source sample are detected in X-ray, with Log $L_{2-10 keV}$ ranging from $\sim$ 40-45 erg/s. 74\% (23 of 31) of our AGN sample and 48\% (31 of 64) of our composite sample is X-ray detected. \cite{Kirkpatrick2012} also report the X-ray detection fractions for the spectroscopically-confirmed AGN sample and find 41$\%$ of Silicate AGN and 92\% of Featureless AGN have an X-ray detection. This characterization is consistent with the idea that dustier nuclear regions, found in Silicate AGN, are more likely to result in X-ray absorption and cause fewer of these objects to be detected in X-ray.

\subsubsection{$L_{X}$ vs. $L_{IR}$: Which sources are under-luminous in X-ray?}

\begin{figure}
\centering
\includegraphics[width=.5\textwidth]{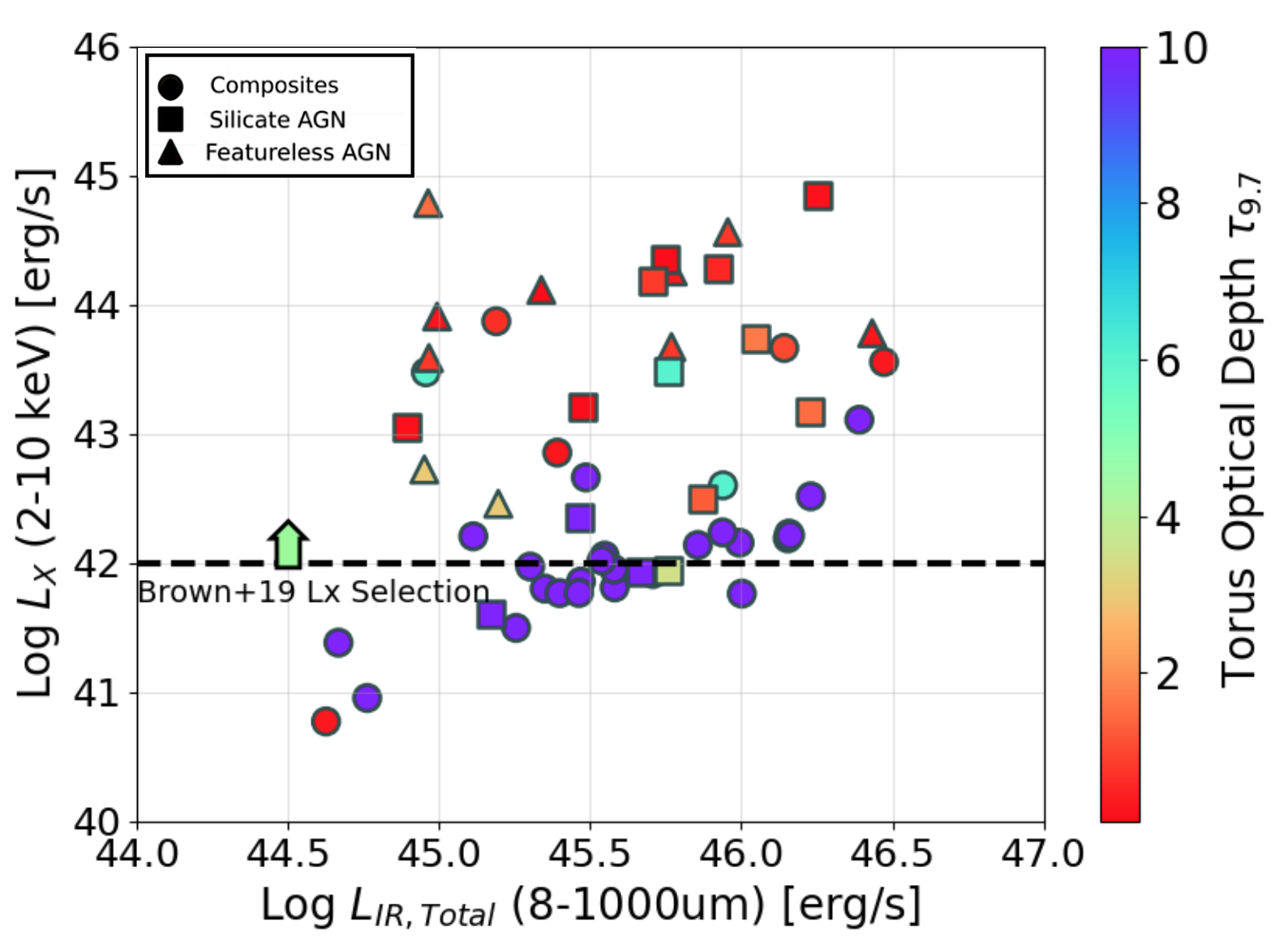}
\caption[4]{\textit{Top:} $L_{X}$ vs. $L_{IR,Total}$ for the 54 objects in our sample with positionally matched X-ray detections \cite{Xue2016,Luo2017}. The colorbar indicates the best-fit torus optical depth $\tau_{9.7}$ from our SED decomposition that employs 100 random samplings of the full AGN torus library \citep{Fritz2006,Feltre2012}.  Featureless, Silicate, and Composite AGN as classified originally in \cite{Kirkpatrick2012} are denoted as triangles, squares, and circles, respectively. The X-ray selection threshold of Log $L_{X} >$ 42 erg/s adopted in \cite{Brown2019} and other works is shown for reference.  }
\label{fig:lx_lir}
\end{figure}

With $L_{X}$ values obtained for nearly half of  our sample, we investigate whether objects deemed optically thick  ($\tau_{9.7} \sim 10$)  with our broadband UV-FIR Decomposition are under-luminous in their X-ray luminosity. In Figure~\ref{fig:lx_lir} we show the X-ray luminosity as a function of total IR luminosity for the X-ray detected sources in our sample color coded by best-fit torus optical depth for the full AGN library run. We see a clear decrease in X-ray luminosity for high optical depth sources with similar IR luminosities, straddling the common X-ray selection threshold $L_{X} \sim 10^{42}$ erg/s. While sources with AGN-dominated MIR SEDs are expected to have high $L_{X}$, we find that both MIR-strong AGN and composite objects fall in this low $L_{X}$ regime when fit with high optical depths.  

A key takeaway from Figure~\ref{fig:lx_lir} is that our SED decomposition method, and its best-fit torus optical depths, are telling a consistent story about the energy reprocessed between X-ray and IR emission by a speculated optically thick nuclear region. Across the full range of $L_{IR}$, objects fit with an optically thick AGN component are consistently under-luminous in X-ray luminosity compared to objects with lower $\tau_{9.7}$.  \citep{Mullaney2010} also address the reprocessed energetics of X-ray and IR luminosity in AGN in the Chandra Deep Field South (CDF-S). They find that the $L_{X}/L_{IR}$ ratio increases as a function of redshift ( $0< z< 2.5$) for low-to-moderate X-ray luminous sources with $L_{X} \sim 10^{42-43}$ erg/s and remains flat for higher $L_{X} \sim 10^{43-44}$ erg/s. Their results imply that low $L_{X}$ objects, where heavy absorption of X-ray AGN photons may occur, become less apparent at high redshift if $L_{X}/L_{IR}$ increases to reveal more X-ray photons. Physically, this implies either a change in dusty torus properties such as increased dust covering factor.  However, our Figure~\ref{fig:lx_lir} suggests that $L_{X}/L_{IR}$ does not evolve with redshift and we may be equally as likely to see diminished X-ray emission, and an optically thick torus, in IR-bright sources in the local universe and the high redshift universe.

\subsubsection{ $L_{AGN,IR}$ vs. $L_{X}$ and Obscuration thresholds }

 \begin{figure*}
\centering
\includegraphics[width=\textwidth]{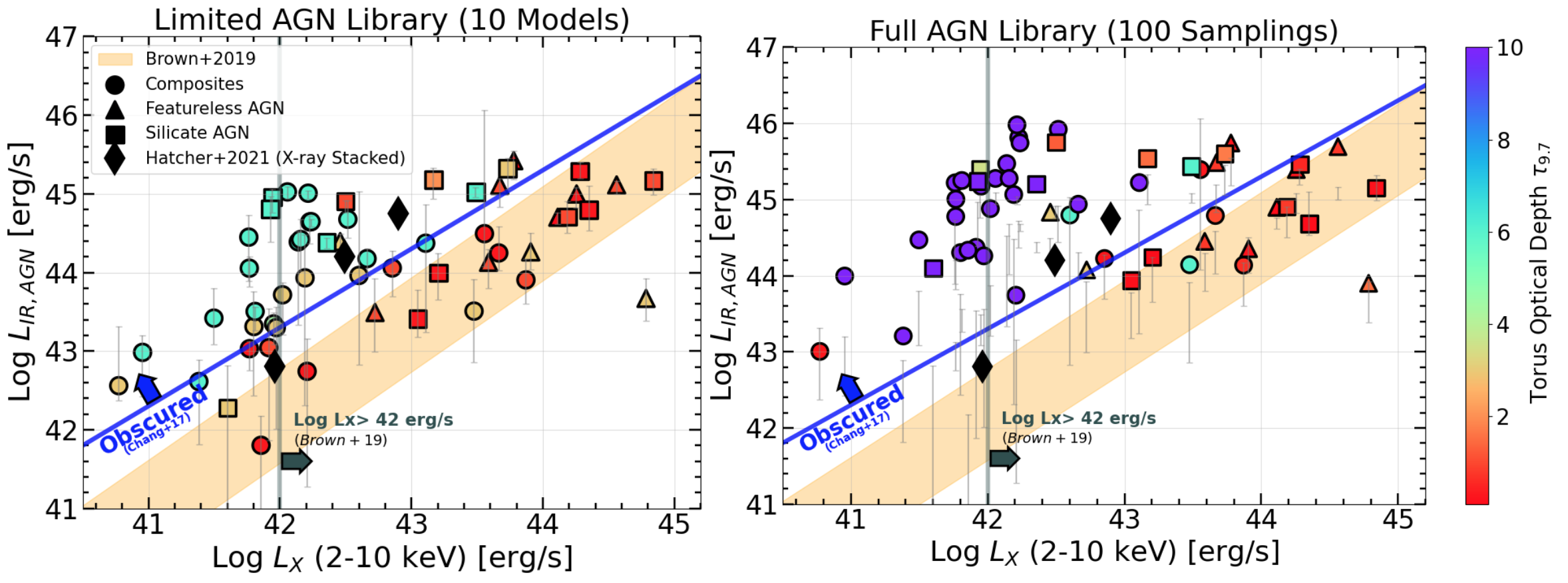}

\caption[3]{$L_{IR,AGN}$ vs 2-10 keV X-ray luminosity for our sample applying both a limited (left panel) and full (right panel) set of AGN templates to broadband SED fitting. AGN luminosity is derived by integrating under the best-fit AGN templates in the rest-frame 8-1000$\mu$m range. Objects are color-coded by respective best- fit $\tau_{9.7}$  and circles represent composites, squares represent represent Silicate AGN, and triangles represent Featureless AGN. For comparison, we show the trend derived in the \cite{Brown2019} study in orange, where IR AGN properties are derived with SED3FIT applying the identical limited set of 10 AGN templates tested in this work. The blue line indicates a derived AGN obscuration threshold in \cite{Chang2017} where objects with $L_{IR,AGN}/L_{X}$ >20 are ``obscured".  Dark gray diamonds represent averaged data from the COSMOS field study \cite{Hatcher2021} for three MIR AGN fraction bins derived via X-ray stacking. $L_{IR,AGN}$ for these data points is derived by extrapolating $f_{FIR,AGN}$ from $f_{MIR,AGN}$ using the empirically-derived trend shown in Figure~\ref{fig:mirfir}. }
\label{fig:liragn}
\end{figure*}

Next we explore the role of torus optical depth in the relation between  $L_{IR,AGN}$ and $L_{X}$ in context with literature trends. This $L_{IR,AGN}$ and $L_{X}$ parameter space can shed light on (1) how cleanly objects separate out into obscuration thresholds based on optical depth and (2) whether a linear trend is the best way describe these quantities for IR-bright objects that may host obscured AGN.  In Figure~\ref{fig:liragn} we show the $L_{IR,AGN}$ - $L_{X}$ relation for our sample color-coded by best-fit torus optical depth. We define $L_{IR,AGN}$ as the AGN luminosity in the rest-frame 8-1000$\mu$m range, derived by integrating under the best-fit AGN template(s) in this range.  We compare our $L_{IR,AGN}$ - $L_{X}$ relation with results from two similar studies by \cite{Brown2019} and \cite{Hatcher2021}. The AGN obscuration threshold estimated in \cite{Chang2017} is also shown; in this work they define any object with $L_{IR,AGN}/L_{X}$ >20 as obscured.

\cite{Brown2019} perform an extensive analysis between X-ray and IR properties for AGN in the Bootes field for X-ray selected AGN with Log $L_{2-10}$ > 42 erg/s. The X-ray Bootes survey is shallower than the X-ray surveys for our  fields, GOODS-N and ECDFS; thus, we match our sample with theirs in $L_{X} - z$ space and identify the subset of matched sources in the comparison. Due to the mismatch in X-ray survey depths, only sources with $L_{X}> 10^{43}$ erg/s overlap with this specific sample selection; however we still show their derived trend extrapolated down to low L$_{X}$ in Figure~\ref{fig:liragn}. To derive IR AGN parameters, \cite{Brown2019} also use SED3FIT and employ the limited set of 10 AGN templates included with the code download. These 10 theoretical AGN templates are identical to the limited library we test throughout this work, allowing for a more direct comparison of IR AGN properties. 

\cite{Hatcher2021} study the relationship between $L_{IR,AGN}$ and $L_{X}$ in the COSMOS field, performing stacking on X-ray undetected sources to estimate X-ray emission. Their $L_{IR,AGN}$ values are estimated using the $f_{AGN,IR}$ vs. $f_{AGN,MIR}$ trend derived in \cite{Kirkpatrick2015} (also shown in Figure~\ref{fig:mirfir}).  For each $f_{AGN,MIR}$ the authors use this trend to compute an $f_{AGN,IR}$ , where MIR AGN fractions are adopted from \cite{Chang2017} MAGPHYS+AGN fits.  As Figure~\ref{fig:mirfir} illustrates, this trend \citep{Kirkpatrick2015} predicts significantly lower FIR AGN contributions than FIR AGN contributions from our full AGN library runs. As a result, although their sample has a lower $L_{IR}$ range, the $L_{IR,AGN}$ values in \cite{Hatcher2021} are significantly lower than those resulting from our optically thick solutions. 


Results from both the limited and full AGN library tell a consistent story with the \cite{Chang2017} obscuration threshold, showing mostly objects with $\tau_{9.7}$ > 6.0 as obscured. The main difference is higher $L_{IR,AGN}$ values in the full AGN library results, as expected given their higher FIR AGN contribution.

To compare with data from the literature, our $L_{IR,AGN}$ results from the limited AGN library runs roughly follow the trend in \cite{Brown2019} for $L_{X}$ > $10^{43}$ erg/s where the sample selections overlap. In this regime, unlike in \cite{Brown2019}, a handful of sources lie within the \cite{Chang2017} obscuration criterion with higher SED3FIT-derived $L_{IR,AGN}$ values pushing them above the line. 

 Our work provides a follow-up to the \cite{Hatcher2021} study by probing the $L_{X}$ $\sim$ 10$^{42}$ erg/s regime with deeper X-ray observations rather than using stacking.  $L_{IR,AGN}$ values in \cite{Hatcher2021} are extrapolated using the empirical $f_{FIR,AGN}$ vs $f_{MIR,AGN}$ relation in \cite{Kirkpatrick2015} and are generally lower than our $L_{IR,AGN}$ values by 1-2 dex in this low-$L_{X}$ regime. Of their stacked data (3 points binned by MIR AGN fraction), only the lowest MIR AGN fraction bin at $L_{X}$ $\sim$ 10$^{42}$ erg/s falls in the optically thin unobscured regime. This is inconsistent with our findings; however, the authors conclude that the galaxies in this bin are likely a mixture of SFGs, low luminosity AGN, and obscured AGN. In this context, this work and the results in Figure~\ref{fig:liragn} confirm that the supermassive black hole activity of objects in this low $L_{X}$ regime is unclear.


 Our $L_{IR,AGN}$ values are computed directly via SED fitting and represent a range of plausible solutions corresponding to different AGN torus models.  We therefore find it is possible for heavily obscured luminous AGNs, with $L_{IR,AGN} \sim 10^{44-46}$ erg/s, to occupy the $L_{X}$ $\sim$ 10$^{42}$ erg/s range. The alternative scenario in which these objects have much lower $L_{IR,AGN}$, as shown in the \cite{Hatcher2021} stacking analysis, demonstrates the sensitivity of this trend and $L_{IR,AGN}$ to decomposition method. These are objects that likely have weak MIR AGN contributions; as a result, they might be commonly excluded from AGN studies and/or mistaken for low-luminosity AGN. \cite{Lambrides2020} propose a similar conclusion via a more detailed X-ray spectral fitting analysis, stating that there may be a large population of obscured AGN with low $L_{X}$ that are disguised as low-luminosity AGN. Overall, the inclusion of these optically thick models shows that $L_{IR,AGN}$ and $L_{X}$ may not follow a clear linear trend, since there may be more obscured AGN than expected to refute the idea that low $L_{X}$ implies low $L_{IR,AGN}$. We also note that within these trends we find no dependence on redshift or stellar mass, and no difference between objects in the GOODS-N vs. ECDFS fields.
 
It remains unclear whether the physical interpretation of these regions as perfectly smooth optically thick dust tori should be accepted at face value (see Section~\ref{subsec:clumpysmooth}). However,  we maintain that our SED fitting paints a reasonable picture of how torus optical depth relates $L_{X}$ and $L_{IR,AGN}$ when X-ray photons may be absorbed by large columns of dust. The frequency at which we should see these optically thick sources in the high redshift universe is also ambiguous; however,  high resolution local observations of sources such as Arp220 support the idea that such galaxies exist nearby and therefore may also exist at high redshifts.  The omnipresence of these objects at high redshift is also supported by estimates of the dust-obscured SFRD and the X-ray background that point to large populations of hidden dusty galaxies and AGN at high redshift.

\section{Summary}
\label{sec:summary}

The objective of this work is to  demonstrate the systematic consequences of decomposing broadband UV/Optical- IR AGN SEDs for sources ranging in AGN strength ($0 \%< f_{MIR,AGN}< 100 \%$ ) and redshift ($0.4 <z < 2.7$) when we fully consider the predicted variability in AGN model parameter space that affects its MIR-FIR SED shape. We use SED3FIT \citep{Berta2013} to decompose the broadband UV-FIR SEDs of 95 galaxies studied in \cite{Kirkpatrick2012} and selected at 24$\mu$m ($S_{24}> 100 \mu$Jy). We fit the sources with two separate samplings of theoretical AGN model parameter space \citep{Fritz2006,Feltre2012}: (1) a restricted set of 10 models with no random sampling, and (2) 100 random samplings of the full 21000-model library. Each fit is repeated 100 times to create a statistically significant set of solutions that fully captures inconsistencies, degeneracies, and systematics affiliated with AGN model variance and/or sensitive mid-IR data gaps. 

Since disentangling AGN properties becomes increasingly difficult when the host galaxy light overwhelms AGN emission, we are particularly interested in how results vary over a dynamic range of MIR AGN strengths. The range of AGN dominance in our sample is supported by isolated spectral MIR decomposition performed in \cite{Kirkpatrick2012}, where we also demonstrate how this method compares with our values derived via UV-FIR broadband SED decomposition. We find that sampling a wider AGN template parameter space in fitting, by allowing models with higher optical depths ($\tau_{9.7,max} = 10$  vs. $\tau_{9.7,max} = 6$), resulted in overall improved fit quality, higher MIR and FIR AGN fractions, higher AGN bolometric luminosities, and higher optical depths.  Our main results are summarized below.
\\

\textbullet{The optical depth of AGN templates used in SED fitting significantly affects best-fit results and can modify the resultant $L_{Bol}$ luminosity functions and black hole accretion rate density (BHARD). We find that increasing AGN model parameter space to higher optical depths in SED fitting resulted in AGN bolometric luminosities greater by as much as $\sim$ 2 dex for composite galaxies with improved fit quality. The inclusion of MIR-weak AGN sources, the consideration of optically thick AGN models, and higher $L_{Bol}$ values increase both the AGN-incidence rate and $L_{Bol}$ luminosity function in IR-selected samples. As a result, the bright end of the $L_{Bol}$  luminosity function is notably affected by adopted AGN templates when these composite sources are included. }\\

\textbullet{ Torus optical depth and SED decomposition method play a significant role in the relationship between $f_{FIR,AGN}$ and $f_{MIR,AGN}$, where optically thick composites with low $f_{MIR,AGN}$ can have very high $f_{FIR,AGN}$. Strong AGN and composites that exceed $f_{IR,AGN}$ estimates in the empirically estimated trend from \cite{Kirkpatrick2015} are generally fit with very high torus optical depths. In the full AGN library fits, composite objects with high $\tau_{9.7} \sim 10$ also had high $f_{AGN,FIR}$ > 40\% while the same objects fit with the limited library had on average $f_{AGN,FIR}$ <20\%.  These results are consistent with those in \cite{Roebuck2016}, supporting the notion that  $f_{AGN,FIR}$ values in composite AGN/SF galaxies may be underestimated by a factor of $\sim$ 3 in \cite{Kirkpatrick2015}, and similar studies based on largely unobscured AGN observations. Such objects with high  $f_{AGN,FIR}$ are also likely optically thick, supported by both our results and the \cite{Roebuck2016} simulations.}  \\

\textbullet{Across all $L_{IR}$ values, objects fit with an optically thick AGN torus component are consistently under luminous in X-ray luminosity compared with optically thin AGN. We investigate the 2-10 keV X-ray luminosities for our sample and find that for objects with similar total IR luminosities, when the best-fit $\tau_{9.7} \sim 10$, its maximum value in the extended library, the X-ray luminosity is lower by $\sim$ 2 orders of magnitude compared with objects best-fit with lower $\tau_{9.7} < 1$. The majority of these optically thick sources are composites; however, the handful of optically thick strong AGN also exhibit this diminished X-ray luminosity, suggesting the absorption of X-ray photons by a thick dusty torus.}\\

\textbullet{ Deep X-ray observations reveal composite objects with low $L_{X}$ $\sim$ 10$^{42}$ erg/s that are optically thick but have high $L_{IR,AGN}$, caused by the reprocessing of X-ray photons into the IR by a thick dusty torus. We compare our results for $L_{IR,AGN}$ vs. $L_{X}$ with those studied in \cite{Brown2019}, \cite{Hatcher2021}, and an obscuration threshold estimated in \cite{Chang2017} where sources are obscured if $L_{IR,AGN}/L_{X}$ >20. Our results occupy a unique region in $L_{IR,AGN}$ vs. $L_{X}$ parameter space where Log $L_{IR,AGN} \sim$ 44-46 erg/s for objects with Log $L_{X}$ $\sim$ 42 erg/s. These objects are best-fit with their respective maximum allowed torus optical depths. Overall, our results suggest that these objects are candidates for obscured or extremely buried AGN  often excluded from X-ray AGN selections and strict IR-selected AGN studies.}\\

\section{Data Availability}
Data used in this work includes multi-wavelength photometry publicly available in \cite{Kirkpatrick2012} and Spitzer Spectroscopy available online through the NASA/IPAC Infrared Science Archive (IRSA).

\bibliographystyle{mnras}
\bibliography{mybib}{}
\bsp	

\appendix
 \begin{figure*}
\centering
\includegraphics[width=\textwidth]{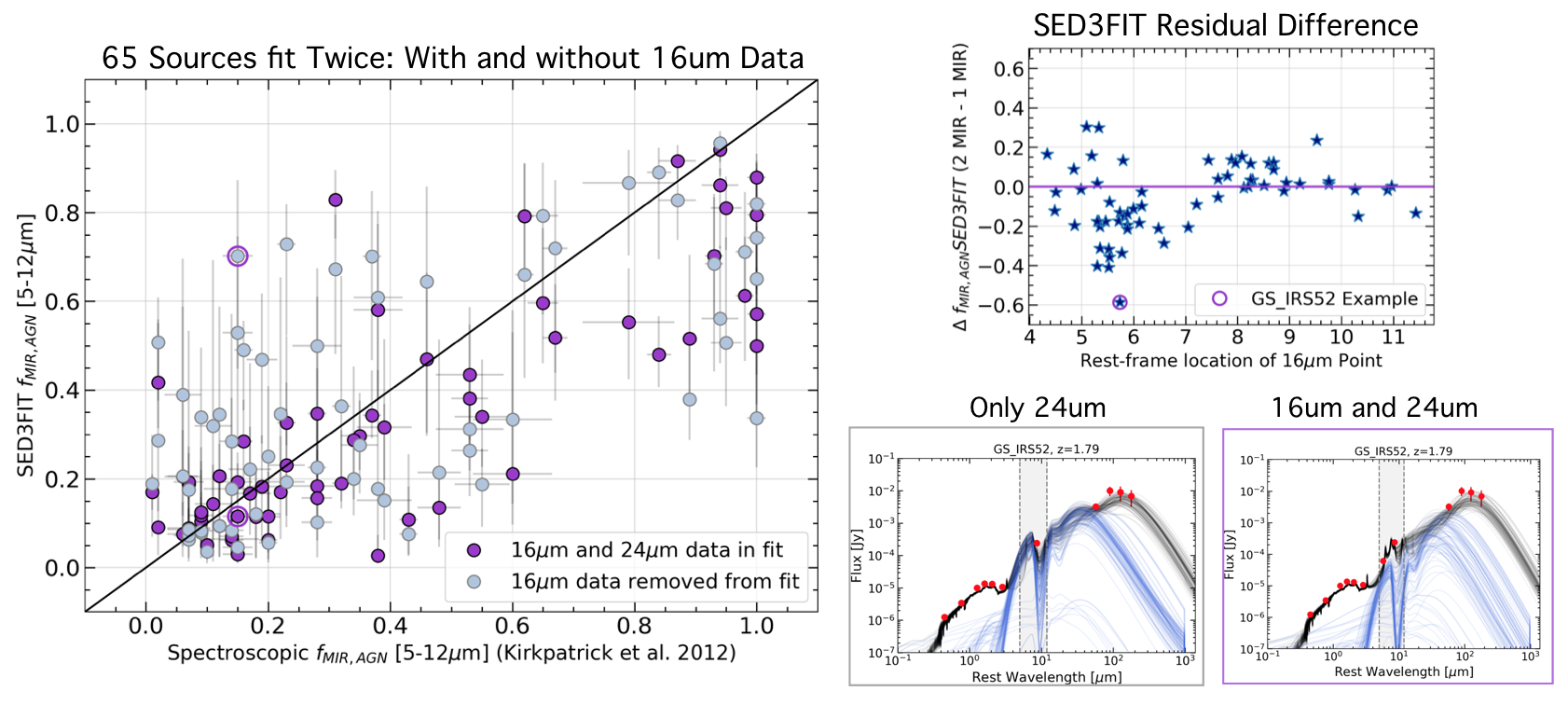}
\caption[1]{A re-visited comparison between MIR AGN fractions derived by decomposing broadband UV-FIR SEDs versus isolated Spitzer/IRS MIR Spectroscopy \citep{Kirkpatrick2012} for 65 sources with both 16$\mu$m and 24$\mu$m data. This comparison is re-visited to directly test the sensitivity of the 16$\mu$m data point on our broadband decomposition. Fits without 16$\mu$m data are shown in gray and fits with 16$\mu$m and 24$\mu$m data are shown in purple. To the right we show the residual difference in \textit{SED3FIT} AGN fractions as a function of the rest-frame location of the 'added' 16$\mu$m point to illustrate how results are more affected when this constraint is added to the rest-frame 5-6$\mu$m range. We show as an example fits to GS\_IRS52 with and without 16$\mu$m data. This composite source had an SED3FIT MIR AGN fraction of $\sim 70 \%$ that decreased to $\sim 15 \%$ when the 16$\mu$m constraint was added, much closer to the one-to-one line comparison. }
\label{fig:no16_fractions}
\end{figure*}

\section{Sensitivity of Results to MIR SED Sampling}

\subsection{Fitting sources with and without 16$\mu$m data}
Differences have been attributed to AGN strength, significant gaps in mid-IR data, and redshift. To finalize this picture, we focus on the mid-IR region in greater detail to discover which constraints produce consistent results versus which constraints, or lackthereof, yield results more subject to scatter. To do so, in this section we experiment with the removal of the 16$\mu$m data point from fitting where applicable. Of the 95 sources total, 65 have both 16$\mu$m and 24$\mu$m detections. For these 65 sources we remove the 16$\mu$m data point and re-run our broadband SED decomposition to more robustly pinpoint and quantify systematic effects.

\subsection{MIR AGN Fraction Comparison}

We revisit the comparison with the spectroscopic MIR AGN fraction \citep{Kirkpatrick2012} in the context of 16$\mu$m constraints. Figure~\ref{fig:bothcompare} shows that the presence of 16$\mu$m data in a subset of sources overall yields better agreement with spectroscopic MIR AGN fractions. Composites with 16$\mu$m + 24$\mu$m  constraints agree with the spectroscopic MIR AGN fraction, while those with only 24$\mu$m return MIR fractions $\sim$ 25\% higher than the spectroscopic values. In Figure~\ref{fig:no16_fractions} we show the MIR AGN fraction comparison with the spectroscopic AGN fractions again for the  65 sources fit with and without 16$\mu$m data for the full AGN library sampling.  Beside this comparison we show the residual difference in SED3FIT mid-IR AGN fraction between runs as a function of the location of the ``added" 16$\mu$m point.  The residual difference is meant to show which sources are most responsible for the increased dispersion i.e. where in the mid-IR SED added data affected the results most.

 This demonstration confirms that a second mid-IR point, 16$\mu$m in this case, only significantly affects mid-IR AGN fraction and its calibration with \cite{Kirkpatrick2012} values when added in the rest-frame 5-6$\mu$m range, corresponding to z$\sim$1.5-2.0.    It is this subset of sources that contribute to any over-estimation of AGN fraction with respect to the spectroscopic decomposition, while lower redshift sources are less likely to see this particular systematic effect. In the higher redshift case, results suggests that if the 24$\mu$m constraint is in close proximity to the 9.7$\mu$m Silicate absorption feature, strong Si absorption models dominate the fit and yield high AGN fractions. When a rest-frame 5-6$\mu$m constraint is added, these models are not permitted to reach such high normalizations or dominate fits, resulting in lower MIR AGN fractions by as much as $\sim$ 50\% as shown in the example in Figure~\ref{fig:no16_fractions}.
 
 For stronger AGN, addition of the 16$\mu$m constraint increased the MIR AGN fraction most significantly near rest-frame $\sim 5 \mu$m by $\sim$ 30\%. This change brought the MIR AGN fraction values closer to the one-to-one line in the comparison, suggesting that the negative $\sim$ 20\% MIR fractional offset between methods for AGN may also be driven by lack of constraints in this narrow range. Since we do not have 16$\mu$m data for the remaining sources, this exercise cannot confidently conclude that the remaining outlying sources should be in closer agreement with the spectroscopic MIR AGN fractions. However, our test suggests that lack of rest-frame $\sim$ 5-6$\mu$m coverage is likely to yield MIR AGN fractions that converge from spectroscopically computed fractions in \cite{Kirkpatrick2012}.
 
 Interestingly, for $z<1$ sources that see the largest negative offset from the one-to-one line overall, adding the 16$\mu$m point to the fit does not considerably affect the resultant MIR AGN fraction. In the following subsection we explore this exercise for AGN bolometric luminosity and discuss these objects, and their MIR constraints, in greater detail. 

\subsection{AGN Bolometric Luminosity}

 \begin{figure*}
\centering
\includegraphics[width=.46\textwidth]{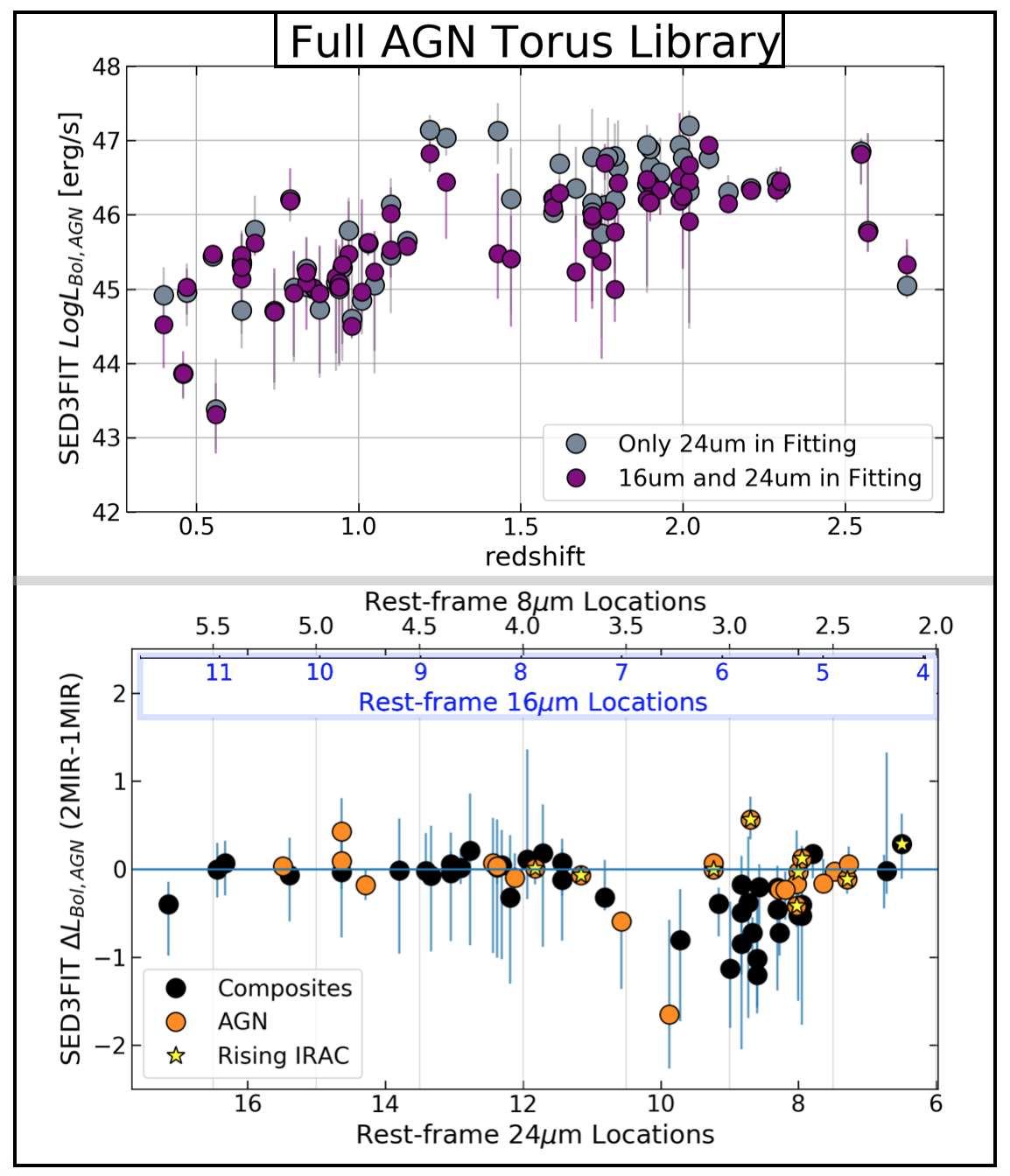}
\includegraphics[width=.45\textwidth]{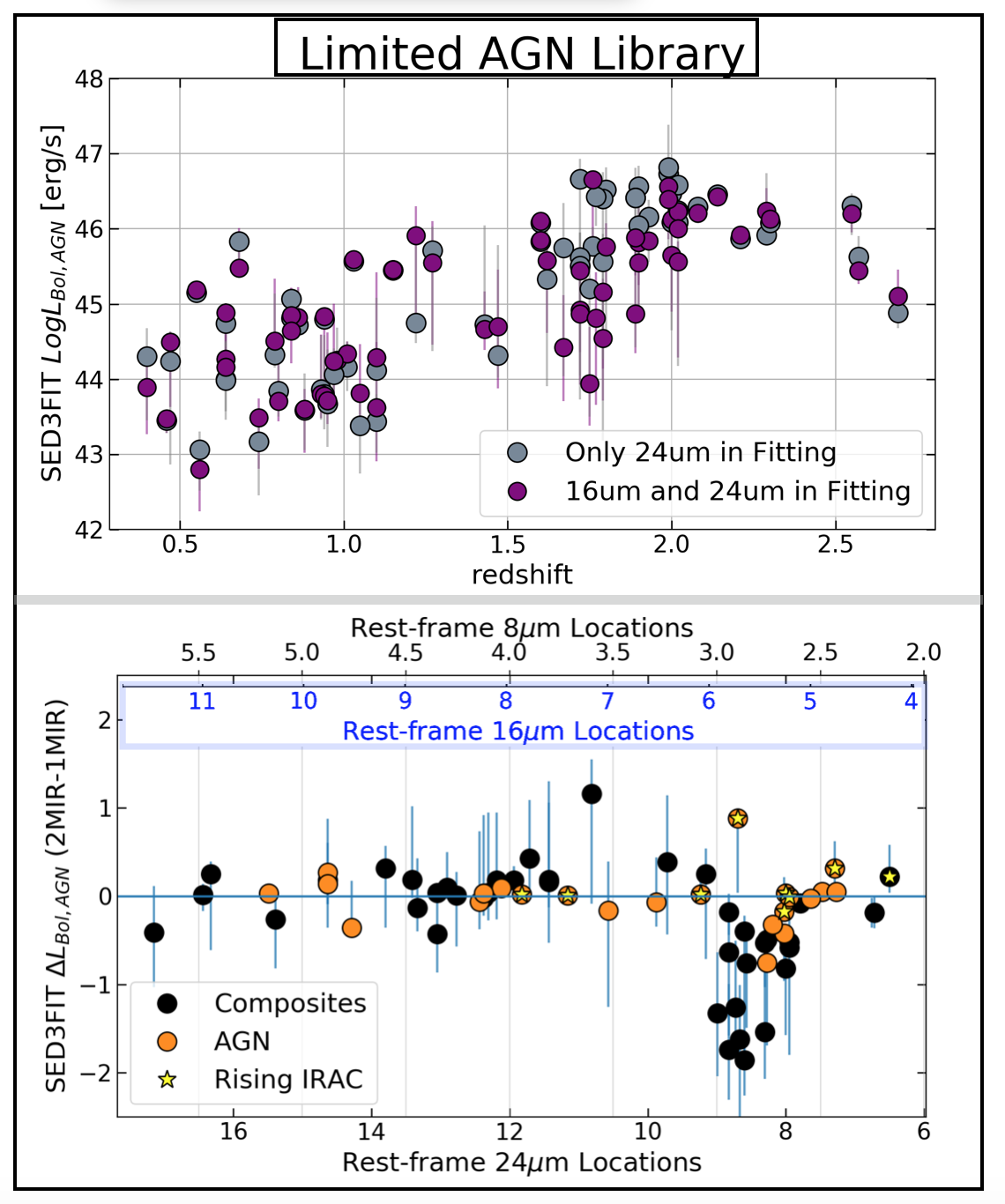}
\caption{Best-fit AGN Bolometric luminosity as a function of redshift for 65 sources fit twice, with and without 16$\mu$m data, for both allowed AGN model settings. The gray points are fits that only include 24$\mu$m while the purple points use both 16 and 24$\mu$m constraints. Below we show the residual difference in derived AGN bolometric luminosity as a function of the rest-frame 8$\mu$m, 24$\mu$m, and ``added" 16$\mu$m point location. AGN are shown in orange, composite sources in black, and sources with a rising IRAC power-law are shown as yellow stars.  Similarly to Figure~\ref{fig:no16_fractions}, these plots reveal that adding this constraint around rest-frame $\sim$ 6$\mu$m affects these quantities most, while the remaining sources see minimal differences. For both AGN model choice settings, lack of coverage in this region can over-estimate AGN luminosity by as much as 2 dex. }
\label{fig:no16_lums}
\end{figure*}

We continue to exclusively work with the subset of 65 sources that has available both 16 and 24$\mu$m data to illustrate systematic behavior of results fit with and without this constraint. Although MIR AGN fractions are affected at a certain redshift range, due to the various FIR shapes of our AGN SEDs it is not trivial whether these constraints also translate strongly to AGN luminosity. For these 65 sources, in Figure~\ref{fig:no16_lums} we show how best-fit AGN bolometric luminosity changes with this constraint removed for both the limited and full set of AGN models. Similarly to Figure\ref{fig:no16_fractions}, we include both fits in the same figure, this time as a function of redshift. Alongside we show the residual difference in fits regarding the presence or absence of 16$\mu$m data as a function of the rest-frame location of 8$\mu$m, 24$\mu$m and 16$\mu$m points.   

Similarly to MIR AGN Fraction, these values only significantly change when this point is added around rest-frame $\sim$ 5-6 $\mu$m. This corresponds to galaxies at $z \sim 1.5-2$.  However, the redshifted locations of the 8$\mu$m point reveals that shorter wavelength constraints may play a significant role in the stability of these fit results. \cite{Henry2019} also find that the systematic effects of AGN contribution via SED decomposition is redshift-dependent. Focusing on the redshifted location of the 24$\mu$m and 70$\mu$m points, they find that z$\sim$3 sources have systematically lower AGN contributions when derived with CIGALE SED fitting vs. with spectroscopy. 

Overall, we can use this information to piece together a cohesive summary of systematics that should be considered for future applications of broadband SED Decomposition. Sources without a rising IRAC power-law and with no coverage in rest-frame 3-8$\mu$m are the most subject to degeneracies, suggesting over-estimated AGN bolometric luminosity. This value is subject to change by as much as $\sim$ 2 dex when a constraint around 6$\mu$m is added.  These sources therefore represent a significant systematic that has the most power to bias results in one direction. On the other hand, the remaining sources with stabler best-fit parameters show the power of constraints in these regions. When an SED has coverage at rest-frame $\sim$ 3-5 $\mu$m, results are less likely to be systematically skewed in a given direction. This also applies for the highest redshift sources in the sample with a constraint at $\sim$ 2$\mu$m and $\sim$ 6$\mu$m. Although these objects have a large gap between 2-6$\mu$m when only 24$\mu$m observations are considered, their 6$\mu$m constraint is enough to leave their results unperturbed by additional constraints in between this gap. This  further supports the narrative that a constraint near $\sim$ 6$\mu$m is necessary for consistency when the 3-5$\mu$m region is unconstrained.  Though these objects may have poorly constrained AGN luminosities, gaps in their mid-IR SED  do not radically affect their fit as much.

\label{lastpage}
\end{document}